\documentclass[a4paper,12pt]{article}
\usepackage{jheppub}
\usepackage[dvipsnames]{xcolor}
\usepackage{tikz}
\usepackage{pgfplots}
\usepackage[T1]{fontenc}
\usepackage[]{slashed}
\usepackage[]{bm}
\usepackage{physics}
\usepackage{dsfont}

\usepackage{graphicx}
\usepackage{epstopdf}
\usepackage{bbm,amsmath,graphicx,amssymb,amsfonts,amsthm}

\def\eps{{\epsilon}}

\def\a{\alpha}
\def\b{\beta}

\newcommand{\Pf}{\text{Pf}\,}

\newcommand{\PT}{\text{PT}}

\newcommand{\f}{\slashed{f}}
\newcommand{\ks}{\slashed{k}}

\newcommand{\dsum}[2]{\sum_{\substack{#1\\#2}}}

\newcommand{\black}[1]{\textcolor{black}{#1}}

\pgfdeclarelayer{bg}    
\pgfsetlayers{bg,main}  

\definecolor{Mathematica}{HTML}{ed192d}

\usepackage[UKenglish]{isodate}
\usepackage{float}

\usetikzlibrary{shapes.misc}
\usetikzlibrary{arrows}
\usetikzlibrary{decorations.markings}
\usetikzlibrary{positioning,fit}
\usetikzlibrary{patterns}

\tikzset{cross/.style={cross out, draw=black, minimum size=2*(#1-\pgflinewidth), inner sep=0pt, outer sep=0pt},
	cross/.default={2pt}}

\newcommand{\polygonn}[2][]{
	\pgfmathsetmacro{\angle}{360/#2}
	\pgfmathsetmacro{\startangle}{-90-\angle/2}
	\pgfmathsetmacro{\y}{cos(\angle/2)}
	\tikzstyle{vertex}=[circle,fill=black,minimum size=7pt,text width = 7pt,inner sep=0pt]
	\foreach \i in {1,2,...,#2} {
		\pgfmathsetmacro{\x}{\startangle - \angle*\i}
		\node[vertex] (p\i) at (\x+\angle:1cm) [label={[label distance= -0.1mm]\x+\angle:$ \i $}] {};
	}

}
\newcommand{\polygonnn}[2][]{
	\pgfmathsetmacro{\angle}{360/#2}
	\pgfmathsetmacro{\startangle}{-90-\angle/2}
	\pgfmathsetmacro{\y}{cos(\angle/2)}
	\tikzstyle{vertex}=[circle,fill=black,minimum size=7pt,text width = 7pt,inner sep=0pt]
	\foreach \i in {1,2,...,#2} {
		\pgfmathsetmacro{\x}{\startangle - \angle*\i}
		\node[vertex,scale=0.1] (\i) at (\x+\angle:1cm) {};
	}
}
\newenvironment{polygon}[1][]
{	\def\myenvargumentII{#1}
	\polygonnn{#1}}
{\polygonn{\myenvargumentII}
}

\newcommand{\one}[1]{\node at (#1,1) [circle,fill=black,inner sep=0pt,minimum size=2mm,label={[yshift={-0.68cm}]\begin{small}$#1$\end{small}}] {};
}

\newcommand{\two}[1]{\node at (#1,2) [circle,fill=black,inner sep=0pt,minimum size=2mm,label={[yshift={-0.1cm}]\begin{small}$#1$\end{small}}] {};
}

\newcommand{\three}[1]{\node at (#1+0.2,2.7) [circle,fill=black,inner sep=0pt,minimum size=2mm,label={[yshift={-0.1cm}]\begin{small}$#1$\end{small}}] {};
}

\newcommand{\dbox}[2]{ \node at (#1,#2) [rectangle,fill=white,draw=black, minimum width=0.6cm,
	minimum height = 0.3cm] {};
\node at (#1,#2) [rectangle,draw=black,pattern=north west lines, pattern color=black, minimum width=0.6cm,
	minimum height = 0.3cm] {};
}
\newcommand{\baselineos}[2]{\node at (#1,0) [circle,fill=black,inner sep=0pt,minimum size=2mm,label=below:\begin{small}$#2$\end{small}] {};}

\newcommand{\oneos}[2]{\node at (#1,1) [circle,fill=black,inner sep=0pt,minimum size=2mm,label={[yshift={-0.68cm}]\begin{small}$#2$\end{small}}] {};
}

\newcommand{\twoos}[2]{\node at (#1,2) [circle,fill=black,inner sep=0pt,minimum size=2mm,label={[yshift={-0.1cm}]\begin{small}$#2$\end{small}}] {};
}

\newcommand{\threes}[1]{\node at (#1,3) [circle,fill=black,inner sep=0pt,minimum size=2mm,label={[yshift={-0.1cm}]\begin{small}$#1$\end{small}}] {};
}

\newcommand{\baseline}[1]{\node at (#1,0) [circle,fill=black,inner sep=0pt,minimum size=2mm,label=below:\begin{small}$#1$\end{small}] {};}

\newcommand{\num}[3]{\begin{pgfonlayer}{bg}\draw[ultra thick,black] (#1,#3) -- (#2,#3);\end{pgfonlayer}}

\newcommand{\numtwo}[3]{\begin{pgfonlayer}{bg}\draw[ultra thick,black,dotted] (#1,#3) -- (#2,#3);\end{pgfonlayer}}

\newcommand{\edgeone}[4]{\begin{pgfonlayer}{bg}\draw[ultra thick,black] (#1,#3) -- (#2,#4);\end{pgfonlayer}}

\newcommand{\edgetwo}[5]{\begin{pgfonlayer}{bg}\draw[ultra thick,black,line cap=round] (#1,#4) -- (#2,#5+0.13) -- (#3,#5+0.13);\end{pgfonlayer}}

\newcommand{\edgesingle}[4]{\begin{pgfonlayer}{bg}\draw[ultra thick,black] (#1+0.2,#3-0.3) -- (#2,#4);\end{pgfonlayer}}

\newcommand{\edgesingletwo}[5]{\begin{pgfonlayer}{bg}\draw[ultra thick,black,line cap=round] (#1+0.2,#4-0.4) -- (#2,#5-0.13) -- (#3,#5-0.13);\end{pgfonlayer}}

\newcommand{\finalone}[1]{\begin{pgfonlayer}{bg}\draw[ultra thick,black] (#1,1) -- (1,0);\end{pgfonlayer}}

\newcommand{\finaltwo}[2]{\begin{pgfonlayer}{bg}\draw[ultra thick,black] (#1,1) -- (#2,0.13) -- (1,0.13);\end{pgfonlayer}}

\newcommand{\finalthree}[1]{\begin{pgfonlayer}{bg}\draw[ultra thick,black] (#1,2) -- (1,0);\end{pgfonlayer}}

\newenvironment{BCJa}[1]{
	\node at (1,0) [circle,,fill=black,inner sep=0pt,minimum size=2mm,label=below:\begin{small}$1$\end{small}]  {};
	\node at (#1,0) [circle,,fill=black,inner sep=0pt,minimum size=2mm,label=below:\begin{small}$#1$\end{small}]
	{};
	\draw[ultra thick,black] (1,0) -- (#1,0);
}

\newenvironment{BCJb}[2]{
	\node at (1,0) [circle,,fill=black,inner sep=0pt,minimum size=2mm,label=below:\begin{small}$1$\end{small}]  {};
	\node at (#1,0) [circle,,fill=black,inner sep=0pt,minimum size=2mm,label=below:\begin{small}$#2$\end{small}]
	{};
	\draw[ultra thick,black] (1,0) -- (#1,0);
}

\newenvironment{BCJc}[2]{
	\node at (1,0) [circle,,fill=black,inner sep=0pt,minimum size=2mm,label=below:\begin{small}$1$\end{small}]  {};
	\node at (#1,0) [circle,,fill=black,inner sep=0pt,minimum size=2mm,label=below:\begin{small}$#2$\end{small}]
	{};
	\draw[ultra thick,black,dotted] (1,0) -- (#1,0);
}

\newenvironment{BCJ}[1][]
{
	\begin{gathered}
	\begin{tikzpicture}[scale=0.57,baseline={([yshift=2ex]current bounding box.center)}]
	\begin{BCJa}{#1}
}
{ 
	\end{BCJa}
	\end{tikzpicture}
	\end{gathered}
}

\newenvironment{BCJ2}[2][]
{
	\begin{gathered}
	\begin{tikzpicture}[scale=0.57,baseline={([yshift=2ex]current bounding box.center)}]
	\begin{BCJb}{#1}{#2}
}
{ 
	\end{BCJb}
	\end{tikzpicture}
	\end{gathered}
}

\newenvironment{BCJ3}[2][]
{
	\begin{gathered}
	\begin{tikzpicture}[scale=0.57,baseline={([yshift=2ex]current bounding box.center)}]
	\begin{BCJc}{#1}{#2}
}
{ 
	\end{BCJc}
	\end{tikzpicture}
	\end{gathered}
}

\title{Scattering of Gravitons and Spinning Massive States from Compact Numerators\!\!\!\!\!\!\!\!\!\!\!\!\!}

\author[a]{N. Emil J. Bjerrum-Bohr,}
\author[a,b]{Taro V. Brown,}
\author[a,c]{Humberto Gomez}

\affiliation[a]{Niels Bohr International Academy, Niels Bohr Institute, University of Copenhagen, Blegdamsvej 17, DK-2100 Copenhagen, Denmark}
\affiliation[b]{Department of Physics, UC Davis, One Shields Avenue, Davis, CA 95616, USA }
\affiliation[c]{Facultad de Ciencias, Basicas Universidad
Santiago de Cali, Calle 5 No 62-00 Barrio Pampalinda Cali, Valle, Colombia}
\emailAdd{bjbohr@nbi.dk,tvbrown@ucdavis.edu,humberto.gomez@nbi.ku.dk\!\!\!\!\!\!\!\!\!\!\!\!\!}

\abstract{We provide a new efficient diagrammatic tool, in the context of the scattering equations, for computation of covariant $D$-dimensional tree-level $n$-point amplitudes with pairs of spinning massive particles using compact exponential numerators. We discuss how this framework allows non-integer spin extensions of recurrence relations for amplitudes developed for integer spin. Our results facilitate the on-going program for generating observables in classical general relativity from on-shell tree amplitudes through the Kawai-Lewellen-Tye relations and generalized unitarity.  }

\begin{document} 
\maketitle
\flushbottom

\section{Introduction}\label{sec:intro}
Extracting classical physics from theories with relativistic quantized gravitons is an important modern concept \cite{Neill:2013wsa,Bjerrum-Bohr:2013bxa,Vaidya:2014kza,Bjerrum-Bohr:2017dxw,Cristofoli:2019neg,Bern:2020gjj}. It relies critically on the identification of long-distance contributions, the treatise of the quantized theory as an effective field theory, and consistent extraction of classical physics from loop amplitudes as pioneered in Refs. \cite{Iwasaki:1971vb}. 
On-shell tree amplitudes are necessary input for employing unitarity, and consequently, relevant $D$-dimensional gluon amplitudes that produce graviton amplitudes using the Kawai-Lewellen-Tye relations \cite{Kawai:1985xq}, was studied in Ref. \cite{Bjerrum-Bohr:2019nws}.\\[5pt]
In this paper, we extend the analysis of Ref. \cite{Bjerrum-Bohr:2019nws} for particles with non-integer spin. Computation of post-Newtonian and post-Minkowskian physics with masses and spin effects have recently been the focus of Refs. \cite{Arkani-Hamed:2017jhn}, and we will build on this. 
In particular the three-point numerators we will introduce are very reminiscent of the amplitude for scattering of massive spin particles and a graviton discussed in Refs. \cite{Guevara:2018wpp,Guevara:2019fsj,Bautista:2019evw}. Another prerequisite is the introduction of fermions in the scattering equations by Refs. \cite{Edison:2020ehu,Edison:2020uzf}, where non-integer spins are included by modifying an algorithm developed in Refs. \cite{Du:2017kpo,Teng:2017tbo}. The algorithm relies on Bern-Carrasco-Johansson color-kinematics master numerators \cite{Bern:2019prr} in a Del-Duca, Dixon, and Maltoni (DDM) half-ladder basis \cite{DelDuca:1999rs}. This basis has a natural extension in the scattering equation framework and allows an expansion of the reduced Pfaffian as noticed by \cite{Cachazo:2013iea} provided by Kawai-Lewellen-Tye orthogonality \cite{BjerrumBohr:2010ta,BjerrumBohr:2010yc,Johnson:2020pny}, %
\begin{equation} \label{eq:pfaff}
\Pf'(\Psi) = \sum_{\beta\in S_{n-2}}N(1,\beta,n)\PT(1,\beta,n)\,.
\end{equation}
This property is very useful and allows for example the application of scattering equation integration rules \cite{Baadsgaard:2015ifa} in the context of theories for gluons \cite{Bjerrum-Bohr:2016juj}.  Tree-level color-kinematic numerators can be computed in various ways, for instance, using the scattering equations framework or through the pure-spinor formalism, \cite{Mafra:2015vca}.
\\[5pt]
Ideally, one would want numerators that are crossing symmetric, but the necessity of defining a reference order implies it is not always the case. In the algorithm of \cite{Teng:2017tbo} the computation of $(n-1)!$ diagrams are required for each numerator. Thus generating all master numerators requires a computation of $(n-1)!\times(n-2)!$ diagrams. Ref. \cite{Edison:2020ehu} presented an improved method by constructing crossing symmetric numerators and thereby directly reducing the computation to $(n-1)!$ diagrams.
 \\[5pt]
We will here, discuss an even further improvement -- a new computational tool for such numerators, which only requires the derivation of $(n-2)!$ diagrams. We will discuss applications of this procedure and also how to extend recurrence relations developed for integer spin to non-integer spin.
We organize the paper as follows. In Section \ref{sec:chy}, we review the scattering equations and the numerator construction. We will then present the motivation for our new diagrammatic rules in Section \ref{sec:num}, which will be extended to seven-point and beyond in Section \ref{sec:seven-point}. We introduce general expressions for numerators for fermions and scalars and gravitons in Section \ref{sec:fermion-scalar}. In Section \ref{factorization}, we discuss the extension of the recurrence relations using numerator expressions. Finally, Section \ref{conclusion}, contains our conclusions. There are three appendices.
\section{Amplitudes from scattering equations}\label{sec:chy}
The scattering equation formalism provides $D$-dimensional scattering amplitudes for a large number of theories, Refs. \cite{Cachazo:2013hca,Cachazo:2013iea,Cachazo:2013gna,Cachazo:2014nsa}, through integration over a moduli space 
\begin{equation} \label{eq:chy-int}
\mathcal{A}(1,\dots,n)=\int \dd \mu_{n} ~\mathcal{I}_L\times \mathcal{I}_R\,.
\end{equation}
Here the left and right integrands $\mathcal{I}_L$ and $\mathcal{I}_R$ depend on the field theory in question, and the measure is 
\begin{equation} \label{eq:chy-measure}
\dd \mu_{n}=z_{rs}^2z_{st}^2z_{tr}^2\prod_{\substack{i=1\\i\neq r,s,t}}\dd z_i\:\delta(S_i)\,.
\end{equation}
The $S_i$'s in the delta-function are the scattering equations and they encode $n$ on-shell momenta $(k^\mu_1,\dots,k^\mu_n)$ in terms of auxiliary variables $z_i\in \mathds{CP}^1$ corresponding to punctures on a Riemann sphere.
\begin{equation}
S_i(z)\equiv\sum_{\substack{j=1\\j\neq i}}\frac{2k_i\cdot k_j}{z_{ij}}=0,~~i=1,...,n \,.
\end{equation}
We can decompose the scattering equation integrands into Parke-Taylor factors,
\begin{equation}
\PT(\alpha(1),\alpha(2),\dots,\alpha(n-1),\alpha(n))\equiv \frac{1}{z_{\alpha(1)\alpha(2)}\cdots z_{\alpha(n-1)\alpha(n)} }\,,
\end{equation}
and reduced Pfaffians defined by
 \begin{equation}
 \Pf'\Psi\equiv\frac{(-1)^{i+j}}{z_{ij}}\Pf \Psi_{ij}\,.
 \end{equation}
Here $\Psi_{ij}$ denotes a reduced matrix obtained by removing the $i^{th}$ and $j^{th}$ row and column from the matrix $\Psi$. The matrix $\Psi$ is a function of external momenta and polarizations (see for instance Ref. \cite{Cachazo:2013hca}). \\[5pt]
The scattering equation formalism can be extended to include massive particles, see Ref. \cite{Dolan:2014ega,Zhou:2020umm}, by revising the scattering equations,
\begin{equation}
S_i(z)=\sum_{\substack{j=1\\j\neq i}}\frac{2k_i\cdot k_j+2\Delta_{ij}}{z_{ij}}=0,~~i=1,...,n\,, \label{eq:scat-eq2}
\end{equation}
where for $n$ massive particles, the $\Delta$ matrix is symmetric and related to the masses of the particles
\begin{equation}\label{eq:delta}
\Delta_{i,j}=\Delta_{j,i},~~~\dsum{i=1}{i\neq j}^{n}\Delta_{i,j}=m_j^2\,.
\end{equation}
For two particles of equal mass $m_1=m_n=m$ it is sufficient to consider $\Delta_{1,n}=\Delta_{n,1}=m^2$ and set the remaining $\Delta$'s to zero. It is easy to show that only the ${ i}^{\rm th}$ and ${ j}^{\rm th}$ rows and columns of $\Psi$ are affected by this, so if we reduce the Pfaffian by those rows and columns, we obtain
\begin{equation}\label{mass-massless}
\frac{(-1)^{i+j}}{z_i-z_j}\Pf_{\text{massive}}[\Psi_{i,j}]=\frac{(-1)^{i+j}}{z_i-z_j}\Pf_{\text{massless}} [\Psi_{i,j}]\,.
\end{equation}
We will use that we can always expand the Pfaffian into kinematic (numerator) parts and Parke-Taylor ($z_i$ dependent) parts. If we consider the reduced Pfaffian $\Pf'(\Psi_{1,n})$ for gluons we have
\begin{equation} \label{eq:ddm-expansion}
\frac{(-1)^{1+n}}{z_1-z_n}
\Pf[\Psi_{1,n}]= \sum_{\beta\in S_ {n-2}}N(1,\beta,n)\, \PT(1,\beta,n)\,,
\end{equation}
where $N(1,\beta,n)$ are numerators. Ref. \cite{Fu:2017uzt} showed that these numerators could be constructed algorithmically and this was developed in Ref. \cite{Teng:2017tbo} into a diagrammatic method. 
This method for writing down algebraic expressions for numerators relies on picking a numerator reference order and writing down combinatorial diagrams such as 
\begin{equation}
\begin{aligned}
\raisebox{-.52\height}{	\includegraphics[]{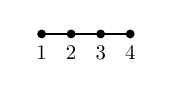}}
\raisebox{-.52\height}{	\includegraphics[]{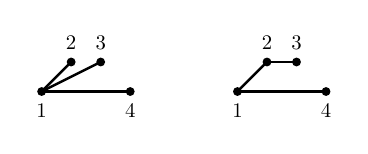}} \quad \ldots
\end{aligned}
\end{equation} 
for which we identify the following algebraic numerator contributions 
\begin{equation} \label{eq:ftr}
(\epsilon_1\cdot f_2 \cdot f_3 \cdot\epsilon_4),~~~~(\epsilon_1\cdot \epsilon_4)(\epsilon_2\cdot k_1)(\epsilon_3\cdot k_1),~~~~(\epsilon_1\cdot \epsilon_4)(\epsilon_3\cdot f_2\cdot k_1)\quad \ldots
\end{equation}
In those expressions, $f_i$ denotes the linearized field strength tensor for particle $i$ with
\begin{equation}
f_{i}^{\mu\nu} \equiv k_i^{\mu}\epsilon^{\nu}_i-k_i^{\nu}\epsilon^{\mu}_i\,,
\end{equation}
polarisation and momentum denoted by $\epsilon_i$ and $k_i$ respectively. We will also use the short-hand notation $\epsilon_i\cdot f_j\cdot \epsilon_k\equiv (\epsilon_{i\,\mu} f_j^{\mu\nu} \epsilon_{k\,\nu})$.\\[5pt]
An important result of Ref. \cite{Edison:2020ehu,Edison:2020uzf} was the extension of the above diagrammatic rules for spinors. We treat these cases through the same diagrams by exchanging the polarization vectors  with spinors and the gluon field strength tensor coming from the baseline by the following spinor replacement
\begin{equation} \label{eq:spinor}
\epsilon_1\to \chi_1,~~~~\epsilon_4\to \xi_4,~~~~f_i^{\mu\nu}\to \slashed{f}_i=\frac{1}{8}f_i^{\mu\nu}[\gamma_\mu,\gamma_\nu].
\end{equation}
Here $\chi_1$ is a ten-dimensional Majorana-Weyl spinor, obeying $\chi_1\ks=0$ and $\xi_n$ is related to $\chi_n$ through $\chi_n=\ks_n\xi_n$. Based on these definitions, we have tested up to six points that the following replacements are valid for $D$-dimensional massive Dirac fermions
\begin{equation} \label{eq:spinor-2}
\epsilon_1\to \bar u_1,~~~~\epsilon_4\to \xi_4,~~~~f_i^{\mu\nu}\to \slashed{f}_i=\frac{1}{8}f_i^{\mu\nu}[\gamma_\mu,\gamma_\nu].
\end{equation}
Here $\bar u_1$ obeys $\bar{u}_1(\ks-m)=0$, and we relate $\xi_n$ to the outgoing spinor, $v_n$ via $v_n= (\ks_n-m)\xi_n$. Thus the fermionic version of the terms \eqref{eq:ftr} are
\begin{equation}
(\bar u_1 \slashed f_2 \slashed f_3 \xi_4),~~~~(\bar u_1 \xi_4)(\epsilon_2\cdot k_1)(\epsilon_3\cdot k_1),~~~~(\bar u_1\xi_4)(\epsilon_3 \cdot f_2 \cdot k_1) \quad \cdots \, 
\end{equation}
Similarly, we have observed that one can construct numerators with particles one and $n$ being scalars by dimensional reduction by setting all products such as
$$
(\epsilon_1 \cdot f_i \cdots f_j\cdot \epsilon_n) \to 0 \, ,
$$
while $\epsilon_1\to 1 $ and $\epsilon_n\to 1$.
Again, considering the expressions \eqref{eq:ftr}, we find the corresponding contributions for two scalars to be
\begin{equation} 
0,~~~~(\epsilon_2\cdot k_1)(\epsilon_3\cdot k_1),~~~~(\epsilon_3\cdot f_2\cdot k_1) \quad \cdots \, 
\end{equation}%
\section{Exponential color-kinematic numerators}\label{sec:num}
We will now deduce a new formalism for numerators to enable even more compact expressions. 
We will denote numerators with a specific reference- and color-ordering, $\rho$ and $\gamma$ respectively, by $N_{ \rho}(1_\sigma,\gamma,n_\sigma)$ where the index $\sigma$ specifies the particle type of legs one and $n$ ($\sigma=0$: spin-zero; $\sigma=1/2$: spin-half and $\sigma=1$: spin-one). Reference symmetric numerators will be defined in the following way
\begin{equation}
N(1_\sigma,\gamma,n_\sigma)=\frac{1}{(n-2)!}\sum_{\rho\in S_{n-2}}N_{\rho}(1_\sigma,\gamma,n_\sigma)\,.
\end{equation}
In the following we will assume that the particles one and $n$ are massive spin-one particles (unspecified $ \sigma$ label indicate that we have taken $ \sigma = 1$). The legs $\{2,3,\dots,(n-1)\}$ are massless gluons. Using the rules established in \cite{Teng:2017tbo} we have, 
\begin{equation} \label{eq:3-point-fei-num}
\begin{aligned}
N_{\{2\}}( 1,2, 3)=&(\epsilon_{2}\cdot k_1)(\epsilon_{1\,\mu}\,\eta^{\mu\nu} \epsilon_{3\,\nu})-(\epsilon_{1\,\mu} f_2^{\mu\nu} \epsilon_{3\,\nu}),
\end{aligned}
\end{equation}
where we have explicitly written out some of the dot products for clarity. We will start with putting this numerator into exponential form by writing it as
\begin{equation}
\begin{aligned}
N( 1,2, 3)&=(\epsilon_2\cdot k_1)\left(\epsilon_{1\,\mu}\! \left[\eta^{\mu\nu}-\frac{f_2^{\mu\nu}}{(\epsilon_2\cdot k_1)}\right]
\epsilon_{3\,\nu}\!\right)=(\epsilon_2\cdot k_1)\left(\!\epsilon_{1\,\mu} \exp[\frac{-f_2}{(\epsilon_2\cdot k_1)}]^{\mu\nu}
	\epsilon_{3\,\nu}\!\right),
\end{aligned}
\end{equation}
since without loss of generality, we can assume that $\epsilon_i\cdot \epsilon_i=0$, $i=2,\ldots, n-1$, so that $f_{i\,\mu\nu}f_i^{\nu\sigma}=0$. \\[5pt]
To extend this formalism beyond three points, we have to perform an average over reference orders. At four points, we have 
\begin{equation} \label{eq:4pt}
\begin{aligned}
N_{\{2,3\}}( 1,2,3, 4)=
&\epsilon_{1\,\mu} f_2^{\mu\nu} f_{3\,\nu\sigma} \epsilon_{4}^{\sigma}-\left( \epsilon_{1\,\mu} f_2^{\mu\nu} \epsilon_{4\,\nu}\right)\left(\epsilon_3\cdot k_{12}\right) -\left( \epsilon_{1,\mu} f_3^{\mu\nu} \epsilon_{4\,\nu}\right)\left(\epsilon_2\cdot k_1\right)\\
&+\left( \epsilon_{1\,\mu} \eta^{\mu\nu}\eta_{\nu\sigma} \epsilon_{4}^{\sigma}\right)\left(\epsilon_2\cdot k_1\right) \left(\epsilon_3\cdot k_{12}\right),\\
\end{aligned}
\end{equation}
where $k_{ij\cdots n}=k_i+k_j+\cdots k_n$. Equation \eqref{eq:4pt} can be written in the suggestive form 
\begin{equation}
\begin{aligned}
N_{\{2,3\}}( 1,2,3, 4)&=(\epsilon_2\cdot k_1)(\epsilon_3\cdot k_{12})\left(\epsilon_{1\,\mu} \left[\eta^{\mu\nu}-\frac{f_2^{\mu\nu}}{(\epsilon_2\cdot k_1)}\right] \left[\eta_{\nu\sigma}-\frac{f_{3\,\nu\sigma}}{(\epsilon_3\cdot k_{12})}\right]
\epsilon_4^{\sigma}\right)\\ &
=(\epsilon_2\cdot k_{1})( \epsilon_3 \cdot k_{12})
\left(
\epsilon_{1\,\mu} \exp[\frac{-f_2}{(\epsilon_{2}\cdot k_1)}]^{\mu\nu} \exp[\frac{-f_{3}}{(\epsilon_3 \cdot k_{12})}]_{\nu\sigma}\epsilon_4^{\sigma}
\right).
\end{aligned}
\end{equation}
Now rewriting the numerator with reference order $(3,2)$, we have,
\begin{equation}
\begin{aligned}
N_{\{3,2\}}( 1,2,3, 4)&=(\epsilon_2\cdot k_{1})( \epsilon_3 \cdot k_{12})
\left(
\epsilon_{1\,\mu} \exp[\frac{-f_2}{(\epsilon_{2}\cdot k_1)}]^{\mu\nu} \exp[\frac{-f_{3}}{(\epsilon_3 \cdot k_{12})}]_{\nu\sigma}\epsilon_4^{\sigma}
\right)\\
&-(k_{1}\cdot k_2)(\epsilon_1\cdot\epsilon_4)(\epsilon_{2}\cdot\epsilon_3),
\end{aligned}
\end{equation}
and we thus arrive at a reference order-averaged numerator at four points,
\begin{equation} 
 \begin{aligned}\label{eq:4ptrule}
 N( 1,2,3, 4)&=(\epsilon_2\cdot k_{1})( \epsilon_3 \cdot k_{12})
 \left(
 \epsilon_{1\,\mu} \exp[\frac{-f_2}{(\epsilon_{2}\cdot k_1)}]^{\mu\nu} \exp[\frac{-f_{3}}{(\epsilon_3 \cdot k_{12})}]_{\nu\sigma}\epsilon_4^{\sigma}
 \right)\\
 &-\frac{1}{2}(k_{1}\cdot k_2)(\epsilon_1\cdot\epsilon_4)(\epsilon_{2}\cdot\epsilon_3).
 \end{aligned}
\end{equation}
At four points, we observe a new feature of numerators, that non-exponential terms start appearing when color and reference order is dissimilar. We will see that this pattern persists to higher point numerators as well. Using the same type of systematics, we can consider the five-point numerator. Repeating the steps we did at four points, we generate the reference ordering crossing symmetric spin-one five-point numerator in the form,
\begin{equation}\label{eq:5ptbch}
\begin{aligned}
& N( 1,2,3,4, 5)=\\ 
&\ (\epsilon_2\cdot k_{1})( \epsilon_3 \cdot k_{12}) (\epsilon_4 \cdot k_{123})
\left(\epsilon_1\cdot \exp[\frac{-f_2}{\epsilon_{2}\cdot k_1}]\cdot \exp[\frac{-f_3}{\epsilon_3 \cdot k_{12}}]\cdot \exp[\frac{-f_4}{\epsilon_4 \cdot k_{123}}]\cdot\epsilon_5\right)\\
&-\frac{1}{2}\left(\epsilon _1\cdot\exp[\frac{-f_2}{\epsilon_2\cdot k_1}]\cdot\epsilon_5\right)(
\epsilon _2\cdot k_1) (\epsilon_3\cdot\epsilon_4)(k_3\cdot k_{12})\\
&-\frac{1}{2}\left(\epsilon_1\cdot \exp[\frac{-f_3}{\epsilon_3\cdot k_{1}}]\cdot\epsilon_5\right)(\epsilon_3\cdot k_{1})(\epsilon_2\cdot\epsilon_4)(k_1\cdot k_2)\\
&-\frac{1}{2}\left(\epsilon _1\cdot\exp[\frac{-f_4}{\epsilon_4\cdot k_{1}}]\cdot\epsilon_5\right)\left(\epsilon _4\cdot k_{1}\right) (\epsilon_2\cdot \epsilon_3)(k_1\cdot k_2)\\
&+\frac{1}{3}(k_1\cdot k_2)(\epsilon_1\cdot\epsilon_5)\left(\left(\epsilon _2\cdot\exp[\frac{-f_3}{\epsilon_3\cdot k_{2}}]\cdot\epsilon_4\right)-(\epsilon_2\cdot \epsilon_3)\left(\epsilon _4\cdot k_{23}\right)\right)\,,
\end{aligned}
\end{equation}
where we for brevity have used dot products between exponentials and polarization vectors which are to be understood in the fashion of equation \eqref{eq:4ptrule}.\\
From the algebraic structure of numerator expressions, a combinatorial algorithm is suggested, which requires the summation of less than $(n-1)!$ diagrams, thus inspired by the diagrammatic notation of Ref. \cite{Edison:2020ehu}, we will now explore this at four, five, and six points.
At four points, by considering the two algebraic contributions, we will now re-derive expression \eqref{eq:4ptrule} by introducing a graphical algorithm. We can define two types of diagrams which interprets the algebraic results as follows. A base-level between points one and four, with points two and three in-between: 
\begin{equation}
\begin{BCJ}[4] 
\baseline{2}
\baseline{3}
\end{BCJ}=(\epsilon_2\cdot k_{1})( \epsilon_3 \cdot k_{12})
\left(
\epsilon_{1}\cdot \exp[\frac{-f_{2}}{\epsilon_{2}\cdot k_1}]\cdot \exp[\frac{-f_{3}}{\epsilon_3 \cdot k_{12}}]\cdot \epsilon_{4}
\right), 
\end{equation}
as well as a diagram with both a base-level between points one and four and a upper level with points two and three connected through the point one:
\begin{equation}
\begin{BCJ}[4]
\one{2}
\one{3}
\num{2}{3}{1}
\finalone{2}
\end{BCJ}=-\frac{1}{2}(k_{1}\cdot k_2)(\epsilon_{2}\cdot\epsilon_3)(\epsilon_1\cdot\epsilon_4)\,.  
\end{equation}
At five points the same logic suggest the following diagrams. Again we start with a base-level between points one and five with points two, three and four in-between:
\begin{equation}
\begin{aligned}
\begin{BCJ}[5]
\baseline{2}
\baseline{3}
\baseline{4}
\end{BCJ}=&(\epsilon_2\cdot k_{1})( \epsilon_3 \cdot k_{12}) (\epsilon_4 \cdot k_{123})\\
&\times\left(\epsilon_1\cdot \exp[\frac{-f_2}{\epsilon_{2}\cdot k_1}]\cdot \exp[\frac{-f_3}{\epsilon_3 \cdot k_{12}}]\cdot \exp[\frac{-f_4}{\epsilon_4 \cdot k_{123}}]\cdot\epsilon_5\right)\,.
\end{aligned}
\end{equation}
Three diagrams with one level where we have the following arrangements of the points two, three and four. Again the lines at the first level are connected through the point one:
\begin{equation}
\begin{aligned}
\begin{BCJ}[5]
\baseline{2}
\one{3}
\one{4}
\finaltwo{3}{2}
\num{4}{3}{1}
\end{BCJ}=&-\frac{1}{2}\left(\epsilon _1\cdot\exp[\frac{-f_2}{\epsilon_2\cdot k_1}]\cdot\epsilon_5\right)(
\epsilon _2\cdot k_1) \black{(\epsilon_3\cdot\epsilon_4)}\black{(k_3\cdot k_{12})}\,,
\\[-5pt]
\begin{BCJ}[5]
\baseline{3}
\one{2}
\one{4}
\finalone{2}
\num{4}{2}{1}
\end{BCJ}=&-\frac{1}{2}\left(\epsilon_1\cdot \exp[\frac{-f_3}{\epsilon_3\cdot k_{1}}]\cdot\epsilon_5\right)(\epsilon_3\cdot k_{1})\black{(\epsilon_2\cdot\epsilon_4)}\black{(k_1\cdot k_2)}\,,
\\[-5pt]
\begin{BCJ}[5]
\baseline{4}
\one{2}
\one{3}
\finalone{2}
\num{3}{2}{1}
\end{BCJ}=&-\frac{1}{2}\left(\epsilon _1\cdot\exp[\frac{-f_4}{\epsilon_4\cdot k_{1}}]\cdot\epsilon_5\right)\left(\epsilon _4\cdot k_{1}\right) \black{(\epsilon_2\cdot \epsilon_3)}\black{(k_1\cdot k_2)}\,.
\end{aligned}
\end{equation}
One diagram with one level where an exponential contribution stemming from the line on the first level is introduced:
\begin{equation}\label{eq:5ptexample}
\begin{aligned}
\begin{BCJ}[5]
\one{4}
\one{2}
\one{3}
\finalone{2}
\num{4}{2}{1}
\end{BCJ}=&\frac{1}{3}(\epsilon_1\cdot\epsilon_5)\black{(k_1\cdot k_2)}(\epsilon_3\cdot k_2)\left(\epsilon_2\cdot \exp[\frac{-f_3}{\epsilon_3\cdot k_2}]\cdot \epsilon_4\right)\,.
\end{aligned}
\end{equation}
Finally, we have a diagram with two levels constructed as follows,
\begin{equation}
\begin{aligned}
\begin{BCJ}[5]
\two{4}
\one{2}
\one{3}
\edgetwo{4}{3}{2}{2}{1}
\finalone{2}
\num{3}{2}{1}
\end{BCJ}=&-\frac{1}{3}(\epsilon_1\cdot\epsilon_5)\black{(\epsilon_2\cdot \epsilon_3)}\black{\left(\epsilon _4\cdot k_{23}\right)}\black{(k_1\cdot k_2)}\,.
\end{aligned}
\end{equation}
Similarly, we can generate the six-point numerator. It contains 25 terms and has a structure reminiscent of the crossing symmetric three-, four- and five-point numerators. For completeness, we have included a full set of the diagrams in Appendix \ref{app:num1}, but we find it illuminating to review a few of them here. One has to consider graphs with six particles with three points on the first level, analogous to these provided by equation \eqref{eq:5ptexample}.
\begin{equation}
\begin{aligned}
\begin{BCJ}[6]
\baseline{2}
\one{3}
\one{4}
\one{5}
\num{3}{5}{1}
\finaltwo{3}{2}
\end{BCJ}&
\\ & \hskip-2.5cm =\frac{1}{3}\black{(k_3\cdot k_{12})}
(\epsilon_4\cdot k_3)\left(\epsilon_3\cdot \exp[\frac{-f_4}{\epsilon_4\cdot k_3}]\cdot \epsilon_5\right)
(\epsilon_{2}\cdot k_1)
\left(
\epsilon_1\cdot \exp[\frac{-f_2}{\epsilon_{2}\cdot k_1}]\cdot \epsilon_6
\right)\,,\end{aligned}\end{equation}\begin{equation}\begin{aligned}
\begin{BCJ}[6]
\baseline{3}
\one{2}
\one{4}
\one{5}
\num{2}{5}{1}
\finalone{2}
\end{BCJ}&
\\ & \hskip-2.5cm=\frac{1}{3}\black{(k_{1}\cdot k_2) }(\epsilon_4\cdot k_2)\left(\epsilon_2\cdot \exp[\frac{-f_4}{\epsilon_4\cdot k_2}]\cdot \epsilon_5\right)(\epsilon_{3}\cdot k_1)
\left(
\epsilon_1\cdot \exp[\frac{-f_3}{\epsilon_{3}\cdot k_1}]\cdot \epsilon_6
\right)\,,\end{aligned}\end{equation}\begin{equation}\begin{aligned}
\begin{BCJ}[6]
\baseline{4}
\one{2}
\one{3}
\one{5}
\num{2}{5}{1}
\finalone{2}
\end{BCJ}&
\\ & \hskip-2.5cm=\frac{1}{3}\black{(k_{1}\cdot k_2)}(\epsilon_3\cdot k_2)\left(\epsilon_2\cdot \exp[\frac{-f_3}{\epsilon_3\cdot k_2}]\cdot \epsilon_5\right)(\epsilon_{4}\cdot k_1)
\left(
\epsilon_1\cdot \exp[\frac{-f_4}{\epsilon_{4}\cdot k_1}]\cdot \epsilon_6
\right)\,,\end{aligned}\end{equation}\begin{equation}\begin{aligned}
\begin{BCJ}[6]
\baseline{5}
\one{2}
\one{3}
\one{4}
\num{2}{4}{1}
\finalone{2}
\end{BCJ}&
\\ & \hskip-2.5cm=\frac{1}{3}\black{(k_{1}\cdot k_2)}
(\epsilon_3\cdot k_2)\left(\epsilon_2\cdot \exp[\frac{-f_3}{\epsilon_3\cdot k_2}]\cdot \epsilon_4\right)
(\epsilon_{5}\cdot k_{1})
\left(
\epsilon_1\cdot \exp[\frac{-f_5}{\epsilon_{5}\cdot k_{1}}]\cdot \epsilon_6
\right)\,.
\end{aligned}
\end{equation}
This composition matches the one explored at lower points.\\[10pt]
\underline{These examples suggest the generic rules for diagrams:}
\begin{itemize}
\item The contributions needed to construct the $n$-point crossing symmetric numerator will involve diagrams with $(n-3)$ upper levels and a base-level. In each diagram, all levels need to have at least two points. An exception is if the single point connects to the end of a level below it.
\item We connect all points in each level horizontally. Next, we draw angled lines to the left connecting all levels. The angled lines connect the first point on a level with the point to the left of it on the level below.
\end{itemize}
\noindent \underline{For each diagram we have to include the contributions according to:}
	\begin{itemize}
		\item Each horizontal line $i<j<k\dots<r<s$, contributes 	\begin{equation}
		\begin{aligned}
		&(\epsilon_j\cdot k_i)(\epsilon_{k}\cdot k_{ij} )\dots(\epsilon_{r}\cdot k_{ijk\dots (r-1)} )\\
		&\left(\epsilon_i\cdot \exp[\frac{-f_{j}}{\epsilon_j\cdot k_i}]\cdot \exp[\frac{-f_{k}}{\epsilon_k\cdot k_{ij}}]\cdots\exp[\frac{-f_{r}}{\epsilon_r\cdot k_{ij\dots(r-1)}}]\cdot \epsilon_s\right)\,.
		\end{aligned}
		\end{equation}
		\item Starting at the first level and working up, remaining lines, $i<j<\dots<r<s$, contribute $(\epsilon_s\cdot k_{ij\dots r})$ unless the $s^{\text{th}}$ point has already been traversed. In that case the contribution is $(k_s\cdot k_{ij\dots r})$.
		\item The overall weight of each coefficient is $\prod_i\frac{1}{\mathcal{T}_i}\frac{1}{\mathcal{B}_i}$ where
		\begin{itemize}
			\item $\mathcal{T}_i$: points in a tree. A tree is a collection of interconnected points that is connected to a point on the base-level.
			\item $\mathcal{B}_i$: points in a branch. A branch is two or more interconnected points, with at least two of them on the same horizontal line, connected to another horizontal line below.
		\end{itemize}
		 \item The overall sign is given by $\prod_{l=1}(-1)^{m_l+1}$, where $m_l$ is number of points in $l^{\text{th}}$ level.
	\end{itemize}
To derive the numerator, we sum all diagrams provided by the rules. 
We have summarised the number of diagrams needed in the computations in Table \ref{tab:n-points}.\\
\begin{table}[htb]
	\begin{center}
		\begin{tabular}{c|c|c|c|c|c|c}
			Number of points & 3 & 4 & 5 & 6 & 7\\
			\hline
			Number of diagrams & 1 & 2 & 6 & 25 & 132
		\end{tabular}
	\end{center}
	\caption{\label{tab:n-points}Number of diagrams needed to compute $n$-point exponential color-kinematics numerators.}
\end{table}\noindent
As observed from the table, the algorithm presented produces numerators with an efficiency of $(n-2)!$ until five points, while for six and seven points, we generate respectively one and twelve additional diagrams. We stress the validity of the rules for any number of points, and will next discuss a further improvement which only generates $(n-2)!$ diagrams for $n$ particles. We achieve this improvement by introducing a notion of off-shell numerators.
\section{Further improvement to the algorithm }\label{sec:seven-point}
To improve the efficiency of the algorithm, we will study additional simplifications of the formalism involving the internal horizontal lines in the diagrams. Let us first note that we can rewrite the expression in \eqref{eq:5ptexample} using the three-point numerator\\[-10pt]
\begin{equation}
\begin{aligned}
\begin{BCJ}[5]
\one{4}
\one{2}
\one{3}
\finalone{2}
\num{4}{2}{1}
\end{BCJ}=&\frac{1}{3}(\epsilon_1\cdot\epsilon_5)\black{(k_1\cdot k_2)}(\epsilon_3\cdot k_2)\left(\epsilon_2\cdot \exp[\frac{-f_3}{\epsilon_3\cdot k_2}]\cdot \epsilon_4\right)\,\\[-5pt]
=&\frac{1}{3}(\epsilon_1\cdot\epsilon_5)\black{(k_1\cdot k_2)}N(2,3,4)\,.
\end{aligned}
\end{equation}\\[-10pt]
This, suggests that a diagrammatic simplification exists where we can build numerators recursively utilizing lower point numerators. For internal lines with three points, this is trivial, so the first non-trivial case is at six points,
\begin{equation} \label{eq:first}
\begin{aligned}
&\begin{BCJ}[6]
\one{2}
\one{3}
\one{4}
\one{5}
\num{2}{5}{1}
\finalone{2}
\end{BCJ}\\
=&\frac{(-1)}{4}\black{(k_1\cdot k_2)}(\epsilon_1\cdot \epsilon_6)(\epsilon_3\cdot k_2)(\epsilon_4\cdot k_{23})\left(\epsilon_2\cdot \exp[\frac{-f_3}{\epsilon_3\cdot k_2}]\cdot \exp[\frac{-f_4}{\epsilon_4\cdot k_{23}}]\cdot \epsilon_5\right)\,,
\end{aligned}
\end{equation}
and
\begin{equation} \label{eq:second}
\begin{aligned}
\begin{BCJ}[6]
\one{2}
\two{3}
\two{4}
\one{5}
\num{3}{4}{2}
\num{2}{5}{1}
\finalthree{3}
\end{BCJ}
=&\frac{1}{8}\black{(k_1\cdot k_2)}(\epsilon_1\cdot \epsilon_6)(\epsilon_2\cdot \epsilon_5)(\epsilon_3\cdot \epsilon_4)(k_3\cdot k_2)\,.
\end{aligned}
\end{equation}
Combining the terms from \eqref{eq:first} and \eqref{eq:second} one finds that sum can be written in the following form
\begin{equation} \label{eq:6ptwith4}
\frac{(-1)}{4}\black{(k_1\cdot k_2)}\black{N(2,3,4,5)}(\epsilon_1\cdot \epsilon_6)\,,
\end{equation}
where $N(2,3,4,5)$ is the four-point numerator obtained from the rules presented previously.
Diagrammatically, we can represent this as
\begin{equation} \label{eq:first1}
\begin{aligned}
&\begin{BCJ}[6]
\one{2}
\one{3}
\one{4}
\one{5}
\num{2}{5}{1}
\finalone{2}
\end{BCJ}+\begin{BCJ}[6]
\one{2}
\two{3}
\two{4}
\one{5}
\num{3}{4}{2}
\num{2}{5}{1}
\finalthree{3}
\end{BCJ}=\begin{BCJ}[6]
\one{2}
\one{3}
\one{4}
\one{5}
\numtwo{2}{5}{1}
\finalone{2}
\end{BCJ}
\end{aligned}
\end{equation}
where the dotted horizontal line marks that the corresponding expression is a lower point numerator. By combining these two diagrams, we end up with 4!=24 diagrams at six points.\\[5pt]
Proceeding to seven-point, we see that this dotted horizontal line has to be considered an off-shell numerator. We consider first the contribution of the following three diagrams given by our rules at seven points, 
\begin{equation} \label{eq:off-shell-2}
\begin{aligned}
&\begin{BCJ}[7]
\one{2}
\one{3}
\one{4}
\one{5}
\num{2}{5}{1}
\two{6}
\edgetwo{6}{5}{2}{2}{1}
\finalone{2}
\end{BCJ}\\
&=\frac{-1}{5}\black{(k_1\cdot k_2)}(\epsilon_1\cdot \epsilon_7)( \epsilon_6\cdot k_{2345})(\epsilon_3\cdot k_2)(\epsilon_4\cdot k_{23})\!\left(\epsilon_2\cdot \exp[\frac{-f_3}{\epsilon_3\cdot k_2}]\cdot \exp[\frac{-f_4}{\epsilon_4\cdot k_{23}}]\cdot \epsilon_5\right)\,,
\end{aligned}
\end{equation}
\begin{equation}\label{eq:sevenpoint1}
\begin{aligned}
&\begin{BCJ}[7]
\one{2}
\two{3}
\two{4}
\one{5}
\num{2}{5}{1}
\num{3}{4}{2}
\edgeone{3}{2}{2}{1}
\threes{6}
\edgetwo{6}{5}{2}{3}{1}
\finalone{2}
\end{BCJ}=\frac{1}{10}(\epsilon_1\cdot \epsilon_7)(k_1\cdot k_2)(\epsilon_6\cdot k_{25})(\epsilon_3\cdot \epsilon_4)(\epsilon_2\cdot \epsilon_5)(k_3\cdot k_2)\,,
\end{aligned}
\end{equation}
\begin{equation}\label{eq:sevenpoint2}
\begin{aligned}
&\begin{BCJ}[7]
\one{2}
\two{3}
\two{4}
\one{5}
\num{2}{5}{1}
\num{3}{4}{2}
\edgeone{3}{2}{2}{1}
\threes{6}
\edgetwo{6}{4}{3}{3}{2}
\finalone{2}
\end{BCJ}=\frac{1}{15}(\epsilon_1\cdot \epsilon_7)(k_1\cdot k_2)(\epsilon_6\cdot k_{34})(\epsilon_3\cdot \epsilon_4)(\epsilon_2\cdot \epsilon_5)(k_3\cdot k_2)\,.
\end{aligned}
\end{equation}
Again we can combine diagrams into a single term. However, unlike the six-point case we cannot recombine them by using the same four-point numerator we used in eq. \ref{eq:6ptwith4}. For the seven-point case we have to introduce an off-shell numerator, since there is a connection to an additional level. Using this off-shell numerator, terms \eqref{eq:off-shell-2}, \eqref{eq:sevenpoint1}, and \eqref{eq:sevenpoint2} can be combined into the simple form
\begin{equation}
\frac{1}{5} (\epsilon_1\cdot \epsilon_7)(k_1\cdot k_2)(\epsilon_6\cdot k_{2345}) N^{[6|5]}(2,3,4,5),
\end{equation}
where the terms in square brackets represent that the sixth particle connects to the fifth leg in the off-shell numerator. The numerator is given by
\begin{equation}\label{eq:off-shell-num-1}
\begin{aligned}
N^{[6|5]}(2,3,4,5)=&
(\epsilon_3\cdot k_{2})( \epsilon_4 \cdot k_{23})
\left(
\epsilon_{2} \cdot \exp[\frac{-f_3}{(\epsilon_{3}\cdot k_1)}] \cdot \exp[\frac{-f_{4}}{(\epsilon_4 \cdot k_{23})}]\cdot \epsilon_5
\right)\\
&-\frac{1}{2}(k_{2}\cdot k_3)(\epsilon_1\cdot\epsilon_s)(\epsilon_{3}\cdot\epsilon_4)\left(\frac{\epsilon_{6}\cdot k_{25}}{\epsilon_{6}\cdot k_{2345}}+\frac{2}{3}\frac{\epsilon_{6}\cdot k_{34}}{\epsilon_{6}\cdot k_{2345}}\right)\,.
\end{aligned}
\end{equation}
Just like before we can represent the terms combined diagrammatically as well
\vskip-0.6cm
\begin{equation}\label{eq:off-shell-1}
\begin{aligned}
&
\begin{BCJ}[7]
\one{2}
\two{3}
\two{4}
\one{5}
\num{2}{5}{1}
\num{3}{4}{2}
\edgeone{3}{2}{2}{1}
\threes{6}
\edgetwo{6}{4}{3}{3}{2}
\finalone{2}
\end{BCJ}
+
\begin{BCJ}[7]
\one{2}
\two{3}
\two{4}
\one{5}
\num{2}{5}{1}
\num{3}{4}{2}
\edgeone{3}{2}{2}{1}
\threes{6}
\edgetwo{6}{5}{2}{3}{1}
\finalone{2}
\end{BCJ}
+
\begin{BCJ}[7]
\one{2}
\one{3}
\one{4}
\one{5}
\num{2}{5}{1}
\two{6}
\edgetwo{6}{5}{2}{2}{1}
\finalone{2}
\end{BCJ}\\
&=
\begin{BCJ}[7]
	\one{2}
	\one{3}
	\one{4}
	\one{5}
	\two{6}
	\edgetwo{6}{5}{2}{2}{1}
	\numtwo{2}{5}{1}
	\finalone{2}
	\end{BCJ}
\end{aligned}
\end{equation}
At seven points we can apply this construction and recombine nineteen diagrams into seven, leaving us with 5!=120 diagrams in total.\\[5pt]
For utilization at higher point orders, we have derived general expressions for generic four-point off-shell numerators. We first consider a branch connected to the first point of an internal four-point numerator.
\begin{equation} 
\begin{BCJ3}[5]{s}
\baseline{1}
\baselineos{3}{q}
\baselineos{4}{r}
\oneos{2}{i}
\finalone{2}
\dbox{6}{1}
\num{2}{6}{1}
\end{BCJ3}=-G \alpha (\epsilon_i\cdot k_1) N^{[i|1]}(1,q,r,s)\,.\label{eq:off-shell-3}
\end{equation}
Here we denote the connecting point $i$, and the box represents the remaining branch with $m$ points. In the above equation, we denote by, $G$, the kinematic information encoded in the branch, $\alpha$ is the overall numerical coefficient of the branch including the line $(1,q,r,s)$, and by $N^{[i|1]}$, we indicate that the $i^{\rm th}$ point of the branch connects to the first point of the numerator line one. 
We decompose \eqref{eq:off-shell-3} in the following two diagrams,
\begin{equation}
\begin{aligned}
\begin{BCJ2}[5]{s}
\baseline{1}
\baselineos{3}{q}
\baselineos{4}{r}
\oneos{2}{i}
\finalone{2}
\dbox{6}{1}
\num{2}{6}{1}
\end{BCJ2}& \\ &\hskip-2.5cm=-G\alpha (\epsilon_i\cdot k_1) (\epsilon_q\cdot k_{1})( \epsilon_r \cdot k_{1q})
\left(
\epsilon_{1}\cdot \exp[\frac{-f_q}{(\epsilon_{q}\cdot k_1)}] \cdot\exp[\frac{-f_{r}}{(\epsilon_r \cdot k_{1q})}]\cdot\epsilon_s
\right)\,,
\\
\begin{BCJ2}[5]{s}
\baseline{1}
\oneos{3}{q}
\oneos{4}{r}
\twoos{2}{i}
\finalthree{2}
\finalone{3}
\dbox{6}{2}
\num{2}{6}{2}
\num{3}{4}{1}
\end{BCJ2}&=\frac{1}{2}G\alpha (\epsilon_i\cdot k_1)(k_{1}\cdot k_q)(\epsilon_1\cdot\epsilon_s)(\epsilon_{q}\cdot\epsilon_r)\,.
\end{aligned}
\end{equation}
Summing the above contributions and comparing with equation \eqref{eq:off-shell-3} we find that the off-shell numerator is 
\begin{equation}
\begin{aligned}
 N^{[i|1]}(1,q,r,s)=&(\epsilon_q\cdot k_{1})( \epsilon_r \cdot k_{1q})
\left(
\epsilon_{1}\cdot \exp[\frac{-f_q}{(\epsilon_{q}\cdot k_1)}] \cdot\exp[\frac{-f_{r}}{(\epsilon_r \cdot k_{1q})}]\cdot\epsilon_s
\right)\\
&-\frac{1}{2}(k_{1}\cdot k_q)(\epsilon_1\cdot\epsilon_s)(\epsilon_{q}\cdot\epsilon_r)\,.
\end{aligned}
\end{equation}
The next case is $q>i>r$ 
\begin{equation} 
\begin{BCJ3}[5]{s}
\baselineos{2}{q}
\baselineos{4}{r}
\oneos{3}{i}
\finaltwo{3}{2}
\dbox{6}{1}
\num{3}{6}{1}
\end{BCJ3}=-G \alpha (\epsilon_i\cdot k_{1q}) N^{[i|q]}(1,q,r,s)\,.
\end{equation}
Denoting the number of particles in the line containing the box and particle $i$ by $m$, this has the following expansion
\begin{equation}
\begin{aligned}
\begin{BCJ2}[5]{s}
\baselineos{2}{q}
\baselineos{4}{r}
\oneos{3}{i}
\finaltwo{3}{2}
\dbox{6}{1}
\num{3}{6}{1}
\end{BCJ2}& \\ &\hskip-2.5cm=-G \alpha (\epsilon_i\cdot k_{1q}) (\epsilon_q\cdot k_{1})( \epsilon_r \cdot k_{1q})
\left(
\epsilon_{1} \cdot \exp[\frac{-f_q}{(\epsilon_{q}\cdot k_1)}] \cdot \exp[\frac{-f_{r}}{(\epsilon_r \cdot k_{1q})}]\cdot\epsilon_s
\right)\,,
\\
\begin{BCJ2}[5]{s}
\baseline{1}
\oneos{2}{q}
\oneos{4}{r}
\twoos{3}{i}
\finalthree{3}
\dbox{6}{2}
\num{3}{6}{2}
\num{2}{4}{1}
\end{BCJ2}=&\frac{1}{2+m}G\alpha (\epsilon_i\cdot k_q)(k_{1}\cdot k_q)(\epsilon_1\cdot\epsilon_s)(\epsilon_{q}\cdot\epsilon_r)\,,
\\
\begin{BCJ2}[5]{s}
\baseline{1}
\twoos{2}{q}
\twoos{4}{r}
\oneos{3}{i}
\finalthree{2}
\finalone{3}
\dbox{6}{1}
\num{3}{6}{1}
\num{2}{4}{2}
\end{BCJ2}=&\frac{1}{2}G\alpha (\epsilon_i\cdot k_1)(k_{1}\cdot k_q)(\epsilon_1\cdot\epsilon_s)(\epsilon_{q}\cdot\epsilon_r)\,,
\end{aligned}
\end{equation}
This, means we can identify,
\begin{equation}
\begin{aligned}
N^{[i|q]}(1,q,r,s)=&
(\epsilon_q\cdot k_{1})( \epsilon_r \cdot k_{1q})
\left(
\epsilon_{1} \cdot \exp[\frac{-f_q}{(\epsilon_{q}\cdot k_1)}] \cdot \exp[\frac{-f_{r}}{(\epsilon_r \cdot k_{1q})}] \cdot\epsilon_s
\right)\\
&-\frac{1}{2}(k_{1}\cdot k_q)(\epsilon_1\cdot\epsilon_s)(\epsilon_{q}\cdot\epsilon_r)\left[\frac{\epsilon_{i}\cdot k_{1}}{\epsilon_{i}\cdot k_{1q}}+\frac{2}{2+m}\frac{\epsilon_{i}\cdot k_{q}}{\epsilon_{i}\cdot k_{1q}}\right]\,.
\end{aligned}
\end{equation}
For the two remaining cases $r<i<s$ and $s<i$ the considerations are similar and we find
\begin{equation}
\begin{aligned}
N^{[i|r]}(1,q,r,s)=&
(\epsilon_q\cdot k_{1})( \epsilon_r \cdot k_{1q})
\left(
\epsilon_{1} \cdot \exp[\frac{-f_q}{(\epsilon_{q}\cdot k_1)}] \cdot \exp[\frac{-f_{r}}{(\epsilon_r \cdot k_{1q})}]\cdot \epsilon_s
\right)\,\\
&-\frac{1}{2}(k_{1}\cdot k_q)(\epsilon_1\cdot\epsilon_s)(\epsilon_{q}\cdot\epsilon_r)\left[\frac{\epsilon_{i}\cdot k_{1}}{\epsilon_{i}\cdot k_{1qr}}+\frac{2}{2+m}\frac{\epsilon_{i}\cdot k_{qr}}{\epsilon_{i}\cdot k_{1qr}}\right]\,,
\end{aligned}
\end{equation}
\begin{equation}
\begin{aligned}
N^{[i|s]}(1,q,r,s)=&
(\epsilon_q\cdot k_{1})( \epsilon_r \cdot k_{1q})
\left(
\epsilon_{1} \cdot \exp[\frac{-f_q}{(\epsilon_{q}\cdot k_1)}] \cdot \exp[\frac{-f_{r}}{(\epsilon_r \cdot k_{1q})}]\cdot \epsilon_s
\right)\\
&-\frac{1}{2}(k_{1}\cdot k_q)(\epsilon_1\cdot\epsilon_s)(\epsilon_{q}\cdot\epsilon_r)\left[\frac{\zeta_{i}\cdot k_{1s}}{\zeta_{i}\cdot k_{1qrs}}+\frac{2}{2+m}\frac{\zeta_{i}\cdot k_{qr}}{\zeta_{i}\cdot k_{1qrs}}\right]\,,
\end{aligned}
\end{equation}
where the vector $\zeta_i$ in the above square bracket, is either a polarization vector $\epsilon_i$ or a momentum vector $k_i$ depending on its connection to the branch.
In these generic expressions we can identify the numerator encountered at seven points from considering equation \eqref{eq:off-shell-num-1} with $i=6$ and $m=1$. These are all the cases at four points for the off-shell numerators. We note that the case $m=0$ corresponds to the on-shell numerator already encountered in \eqref{eq:4ptrule}.\\[2pt]
The procedures presented for the off-shell four-point numerator straightforwardly generalizes to higher point off-shell numerators as well. From five points onwards one has to be mindful of the diagram signs. To illustrate this we can consider the following diagram
\begin{equation} \label{eq:off-shell-51}
\begin{BCJ3}[6]{t}
\baseline{1}
\baselineos{3}{q}
\baselineos{4}{r}
\baselineos{5}{s}
\oneos{2}{i}
\finalone{2}
\dbox{7}{1}
\num{2}{7}{1}
\end{BCJ3}=G \alpha (\zeta_i\cdot k_1) N^{[i|1]}(1,q,r,s,t)\,,
\end{equation}
for which the off-shell numerator is
\begin{equation}
\begin{aligned}
&N^{[i|1]}(1,q,r,s,t) \\
&=(\epsilon_q\cdot k_{1})( \epsilon_r \cdot k_{1q})( \epsilon_s \cdot k_{1qr})
\left(
\epsilon_{1} \cdot \exp[\frac{-f_q}{(\epsilon_{q}\cdot k_1)}] \cdot \exp[\frac{-f_{r}}{(\epsilon_r \cdot k_{1q})}] \cdot \exp[\frac{-f_s}{(\epsilon_s \cdot k_{1qr})}] \cdot \epsilon_{t}
\right)\\
&-\frac{1}{2}(k_{1}\cdot k_q)(\epsilon_{q}\cdot\epsilon_r)(\epsilon_{s}\cdot k_1)
\left(\epsilon_{1} \cdot \exp[\frac{-f_s}{(\epsilon_{s}\cdot k_1)}] \cdot \epsilon_{t}\right)\\
&-\frac{1}{2}(k_{1}\cdot k_q)(\epsilon_{q}\cdot\epsilon_s)(\epsilon_{r}\cdot k_1)
\left(\epsilon_{1} \cdot \exp[\frac{-f_r}{(\epsilon_{r}\cdot k_1)}] \cdot \epsilon_{t}\right)\\
&-\frac{1}{2}( k_r\cdot k_{1q})(\epsilon_{r}\cdot\epsilon_s)(\epsilon_{q}\cdot k_1)
\left(\epsilon_{1} \cdot \exp[\frac{-f_q}{(\epsilon_{q}\cdot k_1)}] \cdot \epsilon_{t}\right)\\
&+\frac{1 }{3}(k_{1}\cdot k_q)\left(\epsilon_{1}\cdot \epsilon_{t}\right)\left[N(q,r,s)-(\epsilon_{q}\cdot\epsilon_r)(\epsilon_s\cdot k_{qr})\right] \left[ (-1)^{3\,\Theta(m-1)} \right]  ,
\end{aligned}
\end{equation}
where $\Theta(x)$ is the standard Heaviside step function, $
\Theta(x) \equiv
0, \,\,\, \text{for} \,\,\, x<0, \ \ 
\Theta(x) \equiv 1, \,\,\, \text{for} \,\,\, x\geq 0. 
$
The sign on the $\frac{1}{3}$ term differs from that of the on-shell numerator because we are considering internal lines and so must include the base-level when determining the sign of each contribution.\\[2pt]
Employing the notion of off-shell numerators, the only modification of the algorithm presented in section \ref{sec:num} is allowing the internal numerators to go off-shell. We have checked the consistency of numerators generated using the above off-shell rules at seven and eight points, with all the seven-point diagrams shown in appendix \ref{app:num2}.
\section{Results for massive fermions, scalars and gravitons} \label{sec:fermion-scalar}
So far, we have discussed results for gluons, and we will now discuss a generalization to Dirac fermions and scalars.
Employing the following replacements, we have checked that one can generate numerators for both integer and non-integer spin,
\begin{equation}\label{Spiners}
\begin{aligned}
&{\sigma}=0:\qquad \epsilon_1\to 1,\,\,~~~~\epsilon_n\to 1,~~\,~~f_i^{\mu\nu}\to 0,\\
&{\sigma}=\frac{1}{2}:\!\qquad \epsilon_1\to \bar u_1,\,~~~\epsilon_n\to \xi_n,~~\,~f_i^{\mu\nu}\to \slashed{f}_i=\frac{1}{8}f_i^{\mu\nu}[\gamma_\mu,\gamma_\nu],\\
&{\sigma}=1:\qquad \epsilon_1\to \epsilon_1,~~~~\epsilon_n\to \epsilon_n,~~~~f_i^{\mu\nu}\to f_i^{\mu\nu},\\
\end{aligned}
\end{equation}
for $k_1^2=k_n^2=m^2$ and provided $\epsilon_i\cdot\epsilon_i=0$. Again we have that $\bar u_1$ obeys $\bar{u}_1(\ks-m)=0$, and we relate $\xi_n$ to the outgoing spinor, $v_n$ via $v_n= (\ks_n-m)\xi_n$.
At three and four points one obtains the following scalar numerators 
\begin{equation} \label{eq:4pt-scalar}
\begin{aligned}
N( 1_0,2, 3_0)=&(\epsilon_2\cdot k_{1}),
\\
N( 1_0,2,3, 4_0)=&(\epsilon_2\cdot k_{1})( \epsilon_3 \cdot k_{12})
-\frac{1}{2}(k_{1}\cdot k_2)(\epsilon_{2}\cdot\epsilon_3),
\end{aligned}
\end{equation}
where the $N( 1_0,3,2, 4_0)$ numerator is obtained from \eqref{eq:4pt-scalar} by exchanging $2\leftrightarrow 3$.
Using this form of numerators, one can verify the results obtained in \cite{Bjerrum-Bohr:2019nws} for the scattering of massive scalars with gluons.
The amplitude calculated using the massive scattering equations is 
\begin{equation}
\begin{aligned}
A( 1_{0},2,3, 4_{0})=&\int \dd \mu_{4} \,  \PT(1,2,3,4) \, \sum_{\beta\in S_2}N( 1_{0},\beta, 4_{0})\PT(1,\beta,4)
\\
=&\frac{(s_{13}-m^2)}{s_{14}}\left[\frac{2(\epsilon_3\cdot k_{4})(\epsilon_3\cdot k_{1})}{(s_{13}-m^2)}+\frac{2(\epsilon_2\cdot k_{1})(\epsilon_3\cdot k_{4})}{(s_{12}-m^2)}+(\epsilon_{2}\cdot \epsilon_3)\right],
\end{aligned}
\end{equation}
where $s_{ij} \equiv (k_i + k_j)^2$.
In the previous sections, we have provided a compact framework for computing scattering amplitudes of two massive vector fields and $(n-2)$ gluons. Using the scattering equations the explicit form of the scattering amplitude \eqref{eq:chy-int} for $n$ gravitons is
\cite{Cachazo:2013gna,Cachazo:2013iea},
\begin{eqnarray}\label{ChyPrescriptionG}
{\cal M}_n = \int d\mu_n \,\, {\rm Pf}^\prime \Psi_n \times {\rm Pf}^\prime \tilde \Psi_n \, ,
\end{eqnarray}
where we identify gravitons by $h_a^{\mu\nu}\equiv \eps^\mu_{a}\tilde\eps^\nu_{a}$, and we have used a $\sim$ to indicate that there is no need to use the same reduced Pfaffian in the left and right integrands. One can expand this Pfaffian in terms of the numerators using \eqref{eq:ddm-expansion}.\\[5pt]
Replacing the vector numerators in this with either scalar or fermion numerators, one can then obtain a plethora of amplitudes with mixed particles.
A convenient route is employing the Kawai-Lewellen-Tye \cite{Kawai:1985xq} relations using the compact formulation through the momentum kernel \cite{BjerrumBohr:2010yc}, which in the scalar case is
\begin{eqnarray}\label{Mn2p2ha}
\!\!\!\!\!\!\!\!\!\!{\cal M} _n ({1}_0,2_h,\dots(n-1)_h,{n}_0)\! =(-1)^{n-3} \sum_{\a \in S_{n\!-\!3} \atop \b \in S_{n\!-\!3}} \!\!\! A_n( { 1}_{0}, \a, (n\!-\!1),{n}_0) \times \! {\cal S}{[\a|\b]_{k_1}} \! \times \! \nonumber\\ \vspace{2.5cm} A_n((n\!-\!1),{ n}_0, \b,{ 1}_{0})\, .
\end{eqnarray}
Here the label $h$ refers to gravitons, and $A_n$ is a color-ordered amplitude of two massive scalars and $(n-2)$ gluons. The momentum kernel ${\cal S}[\alpha|\beta] $ is defined by
\begin{equation}
{\cal S}[i_1,\ldots,i_k | j_1,\ldots, j_k]_{k_1} \equiv \prod_{t=1}^k\left(\tilde s_{{i_t} 1}+\sum_{q>t}^k \Theta(i_t,i_q)\tilde s_{{i_t}\,{i_q}}\right).\end{equation}
In the above equation $\Theta(i_t,i_q)$ equals zero for identical ordering of the sets ${i_1,...,i_k}$ and ${j_1,...,j_k}$, and is one if the ordering of the legs $i_t$ and $i_q$ is opposite. We have $\tilde s_{ij} \equiv 2 (k_i \cdot k_j)$.
For the three-point amplitude scalar-scalar-graviton amplitude, we immediately get 
\begin{equation}
\begin{aligned}
\mathcal{M}(1_1,2_h , 3_1)
&=A( 1_0,2, 3_0)\times A( 1_1,2, 3_1)=-(\epsilon_2\cdot k_1)^2\left(\epsilon_1 \cdot \exp[\frac{- f_{2}}{(\epsilon_2\cdot k_1)}] \cdot \epsilon_3\right).
\end{aligned}
\end{equation}
Similarly the fermion-fermion-graviton amplitude is 
\begin{equation}
\begin{aligned}
\mathcal{M}( 1_{1/2},2_h , 3_{1/2})
&=A(1_0,2, 3_0)\times A( 1_{1/2},2, 3_{1/2})=-(\epsilon_2\cdot k_1)^2\left(\bar u_1 \exp[\frac{- \slashed{f}_{2}}{(\epsilon_2\cdot k_1)}] \xi_3\right).
\end{aligned}
\end{equation}
This exactly matches the covariant result from \cite{Guevara:2018wpp,Arkani-Hamed:2017jhn} which is calculated in the minimal coupling framework.
The four-point results for scalars and gravitons are 
\begin{equation}
\begin{aligned}
\mathcal{M}( 1_0,2_h,3_h, 4_0)=&A( 1_0,2,3, 4_0)\times {\cal S}{[3|2]}\times A( 1_0,3,2, 4_0)\\
&\!\!\!\!\!\!\!\!\!\!\!\!\!\!\!\!\!\!\!\!\!\!\!\!\!\!\!\!\!\!\!\!\!\!\!\!\!\!\!\!\!\!\!\!\!=\frac{(s_{13}-m^2)(s_{12}-m^2)}{s_{14}}\left[\frac{2(\epsilon_3\cdot k_{4})(\epsilon_3\cdot k_{1})}{(s_{13}-m^2)}+\frac{2(\epsilon_2\cdot k_{1})(\epsilon_3\cdot k_{4})}{(s_{12}-m^2)}+(\epsilon_{2}\cdot \epsilon_3)\right]^2,
\end{aligned}
\end{equation}
where ${\cal S}{[3|2]} = s_{14}$. The fermion-gluon amplitude for four particles is obtained as follows
\begin{equation}
\begin{aligned}
A( 1_{1/2},2,3, 4_{1/2})=&\frac{N( 1_{1/2},2,3,4_{1/2})}{s_{12}-m^2}+\frac{N( 1_{1/2},2,3, 4_{1/2}) - N( 1_{1/2},3,2, 4_{1/2})}{s_{14}}
\\
&\!\!\!\!\!\!\!\!\!\!\!\!\!\!\!\!\!\!\!\!\!\!\!\!\!\!\!\!\!\!\!\!\!\!\!\!\!\!\!\!\!\!\!\!\!=
\left(\frac{1}{s_{12}-m^2}+\frac{1}{s_{14}}\right)
(\epsilon_2\cdot k_{1})( \epsilon_3 \cdot k_{12})
\left(
\bar u_{1}\exp[\frac{-\slashed f_{2}}{\epsilon_{2}\cdot k_1}]\exp[\frac{-\slashed f_{3}}{\epsilon_3 \cdot k_{12}}] \xi_{4}
\right) \\
&\!\!\!\!\!\!\!\!\!\!\!\!\!\!\!\!\!\!\!\!\!\!\!\!\!\!\!\!\!\!\!\!\!\!\!\!\!\!\!\!\!\!\!\!\!
-\frac{1}{s_{14}}
(\epsilon_3\cdot k_{1})( \epsilon_2 \cdot k_{13})
\left(
\bar u_{1} \exp[\frac{-\slashed f_{3}}{\epsilon_{3}\cdot k_1}] \exp[\frac{-\slashed f_{2}}{\epsilon_2 \cdot k_{13}}]  \xi_{4}
\right) \\
&\!\!\!\!\!\!\!\!\!\!\!\!\!\!\!\!\!\!\!\!\!\!\!\!\!\!\!\!\!\!\!\!\!\!\!\!\!\!\!\!\!\!\!\!\! -\left(\frac{1}{s_{12}-m^2}+\frac{1}{s_{14}}\right) \frac{1}{2} (k_{1}\cdot k_2)(\epsilon_{2}\cdot\epsilon_3)(\bar u_1\xi_4) +  \frac{1}{2\, s_{14}} (k_{1}\cdot k_3)(\epsilon_{2}\cdot\epsilon_3)(\bar u_1\xi_4)
\\
&\!\!\!\!\!\!\!\!\!\!\!\!\!\!\!\!\!\!\!\!\!\!\!\!\!\!\!\!\!\!\!\!\!\!\!\!\!\!\!\!\!\!\!\!\!=
\frac{ - \bar{u}_1 \slashed{\epsilon}_{2} ( \slashed{k}_{12}+m) \slashed{\epsilon}_{3}  v_4 }{ 4 (s_{12}-m^2)  } +
\frac{ - \bar{u}_1 \slashed{\epsilon}_{2} ( \slashed{k}_{12}+m) \slashed{\epsilon}_{3}  v_4 + \bar{u}_1 \slashed{\epsilon}_{3} ( \slashed{k}_{13}+m) \slashed{\epsilon}_{2}  v_4 }{ 4 \, s_{14}  } \\
&\!\!\!\!\!\!\!\!\!\!\!\!\!\!\!\!\!\!\!\!\!\!\!\!\!\!\!\!\!\!\!\!\!\!\!\!\!\!\!\!\!\!\!\!\!=
\frac{ - 1 }{ 4 (s_{12}-m^2)  }\left( \bar{u}_1 \slashed{\epsilon}_{2} ( \slashed{k}_{12}+m) \slashed{\epsilon}_{3}  v_4  \right) +
\frac{ 1 }{ 4 \, s_{14}}  \left( (\epsilon_2\cdot \epsilon_3) (\bar{u}_1( \slashed{k}_{2}-\slashed{k}_{3}) v_4) \right. \\
&
\qquad\qquad\qquad\qquad\qquad\quad
\left.
- 2 (\epsilon_3\cdot k_2 ) (\bar{u}_1 \slashed{\epsilon}_{2} v_4)  + 2 (\epsilon_2\cdot k_3 ) (\bar{u}_1 \slashed{\epsilon}_{3} v_4) \right)  
\end{aligned}
\end{equation}%
We can use this result to obtain the massive fermion-graviton amplitude, see instance \cite{Bjerrum-Bohr:2013bxa}. (For different extensions and applications of the KLT relation in the context of mixed gauge theory amplitudes, see for instance \cite{KLText}.) It is clear that the efficiency of the algorithm, as well as its properties under different spin configurations, makes it is a powerful tool to compute amplitudes involving graviton interactions and various pairs of massive external particles. 
\section{Recurrence relations with massive spinning particles-version}\label{factorization}
As an alternative to the diagrammatic rules and to gain more insight into the structure of amplitudes, we will now discuss an extension of the recurrence formula obtained in \cite{Bjerrum-Bohr:2018lpz,Gomez:2018cqg,Gomez:2016bmv} (for formulas derived from on-shell recursion, see \cite{Badger:2005zh}). We found the recurrence formula by studying the double-cover approach and the properties of the reduced Pfaffian for gluons. Refs. \cite{Bjerrum-Bohr:2019nws,Zhou:2020mvz} discussed the extension for two massive scalars. Now using our new color-kinematics master numerators for fermions, we will demonstrate that the double-cover recurrence formula is generalizable to scattering amplitudes of two massive fermions and $(n-2)$ gluons.
\\[5pt] 
We start by recalling the recurrence formula for massless gluons of Ref. \cite{Bjerrum-Bohr:2018lpz} 
\begin{align}\label{Gen-C}
	&
	A_n ( 1,2,3,\ldots , n)   
	=  \nonumber\\
	&
	(-1) \sum_M \frac{ A_{n-1} (\bm{1}, n,\ldots , 4, \bm{k}^M_{23} ) \, A_{3}( \bm{k}_{4:1}^M,2,\bm{3}) }{ k_{4:1}^2 }  
	-\sum_{i=1}^{n-3} 
	\left( \delta_{n, 2 \mathbb{N}} + (-)^i \, \delta_{n, 2 \mathbb{N}+1} \right)
	\nonumber \\
	&
	\left[
	\sum_M 
	\frac{A_{n-i}(\bm{1},2, \bm{k}^M_{3:i+3} ,i+4,\dots, n) \, A_{i+2}( \bm{k}^M_{i+4:2}, i+3,\ldots, 4,\bm{3} ) }{ k_{3:i+3}^2 }  
	\right.
	 \\
	&
	-
	2\,
	\left. 
	\sum_L  
	\frac{ A_{n-i}(\bm{1},3, \bm{k}^L_{24:i+3},i+4,\dots, n) \, A_{i+2}( \bm{k}^L_{i+4:13}, i+3,i+2,\ldots, 4,\bm{2} ) }{ k_{24:i+3}^2 } 
	\right].
	\nonumber
\end{align}
where $\bm{ bold}$ labels denote rows/columns removed from the Pfaffian, and we employ 
\begin{align}
	k_{i:j} \equiv k_{i}+k_{i+1}+\cdots +k_{j-1}+k_{j} \,\,\, \, \text{modulo} \ n, \nonumber 
\end{align}\\[-35pt]
\begin{align}\label{gluingR}
	\sum_M \epsilon^{M, \mu}_{k_{i} } \epsilon^{M, \nu}_{k_{j}} \equiv \eta^{\mu\nu} \,, \qquad 
	\sum_L \epsilon^{L, \mu}_{k_i } \epsilon^{L, \nu}_{k_j} \equiv \frac{k^\mu_{i} \, k_{j}^\nu  }{ k_i\cdot k_j }\, , \qquad \text{with}\,\,\,k_i+k_j=0.
\end{align}
The recurrence formula \eqref{Gen-C} was obtained by removing the rows/columns one and three in the reduced Pfaffian and fixing the punctures, $\{ z_1,z_2,z_3,z_4 \}$. This leads to following $\Delta_{p,q}$ parameters for the currents (sub-amplitudes) following the conventions in eq. \eqref{eq:scat-eq2},
\begin{equation}\label{Delta1}
 \Delta_{i,j}\!= \frac{ -K^2 }{2}, \qquad 
\quad \Delta_{K,i}\!= \frac{ K^2 }{2}, \qquad \Delta_{K,j}\!= \frac{ K^2 }{2} ,
\end{equation}
with $i,j\in\{1,2,3,4\}$ and where $K$ is the massive momentum vector, for example, $K=k_{23}$ in $A_{n-1} (\bm{1}, n,\ldots , 4, \bm{k}^M_{23} )$.\\[5pt]
To generalize this formula to include massive non-integer spinning particles, we will employ the
numerator representation of the Pfaffian. For instance, 
since we removed rows/columns one and three in the Pfaffian then, by \eqref{mass-massless} and \eqref{eq:ddm-expansion}
\begin{equation}\label{Pfaff:N}
\frac{ 1}{ z_{1}-z_3 } {\rm \Pf} [ \Psi_{13} ] \, = \,\sum_{\gamma\in S_{n-2} } N(1, \gamma , 3) \, {\rm PT} (1, \gamma, 3),
\qquad 
\end{equation}
the legs one and three can be massive non-integer spinning particles, leading to an amplitude through, 
\begin{align}
	A_n ( {1}_\sigma,2,{3}_\sigma,\ldots , n)   
	\equiv \int d\mu_n\, {\rm PT}(1,2,\ldots, n)\times\!\!\! \sum_{\gamma\in S_{n-2}} N(1_\sigma,\gamma,3_\sigma) \, {\rm PT}(1,\gamma,3),
\end{align}
with $k_1^2=k_3^2=m^2$ and where $\sigma$ denotes the spin. However, the equality \eqref{Pfaff:N} can only be used when the mass of the two massive external legs are identical. For the contribution, $A_{n-1} (\bm{1}, n,\ldots , 4, \bm{k}^M_{23} )$, one has for instance, $k_1^2=m^2$ while $k_{23}^2 = 2k_2\cdot k_3 + m^2$,  
therefore we need to generalize \eqref{Pfaff:N} in such cases. We can achieve this by expressing the reduced Pfaffians of these contributions in terms of shifted color-kinematic master numerators. We employ ${\rm SL}(2,\mathbb{C})$ symmetry preserving shifts, $k_a\cdot k_b \, \rightarrow \, k_a\cdot k_b + \Delta_{a,b}$ in the numerators
\begin{equation}
N(i,\gamma, j) \,\, \rightarrow \, \, N^\Delta(i,\gamma, j),
\end{equation}
where  
\begin{equation}\label{DeltaN}
N^\Delta (i,\gamma_1,....,\gamma_n,j) \equiv N (i,\gamma_1,....,\gamma_n,j) \Big|_{ (k_a\cdot k_b) \rightarrow (k_a\cdot k_b) + \Delta_{a,b} } .
\end{equation}
We now see that when $k_i^2=k_j^2=m^2$, one has,
\begin{equation}\label{NdeltaN}
N(i,\gamma, j) = N^\Delta(i,\gamma, j),
\end{equation}
since the only non-vanishing parameters are $\Delta_{i,j}= \Delta_{j,i}=m^2$. (In the particular case of three-point, the identity \eqref{NdeltaN} is always satisfied). 
This leads to the generalized identity,
\begin{equation}\label{Pfaff:N2}
\frac{ (-1)^{i+j}}{ z_{i}-z_j } {\rm \Pf} [ \Psi_{ij} ] = \sum_{\gamma\in S_{n-2} } N^\Delta(i, \gamma , j) \, {\rm PT} (i, \gamma, j),
\quad \text{with} \, \,\,\, k_i^2\neq k_j^2\neq 0 ,
\end{equation}
which holds on the support of the scattering equations \eqref{eq:scat-eq2}. This identity can be understood from dimensional reduction. It has been checked numerically until six points.  \\[5pt] 
With this machinery in place, deriving non-longitudinal contributions for amplitudes with massive spinning particles is straightforward.
The contributions are given by the expressions
\begin{align}\label{}
	&
	\sum_M \frac{ A_{n-1} (\bm{1}_\sigma, n,\ldots , 4, \bm{k}^M_{23,\sigma} ) \, A_{3}( \bm{k}_{4:1,_\sigma}^M,2,\bm{3}_\sigma ) }{ k_{4:1}^2 -m^2 } ,
	\label{G-nonL1} \\
	&
	\sum_M 
	\frac{ A_{n-i}(\bm{1}_\sigma,2, \bm{k}^M_{3:i+3,\sigma} ,i+4,\dots, n) \, A_{i+2}( \bm{k}^M_{i+4:2,\sigma}, i+3,\ldots, 4,\bm{3}_\sigma ) }{ k_{3:i+3}^2 - m^2 },
	\label{G-nonL2}
\end{align}
where we employ the sum rules for the various spins
{\begin{align}\label{G-glue2}
	\begin{matrix}
		\!\!\!\!\!\!\!\!\text{Spin-one} & \!\!\!\!\!\!\!\!\!\text{Spin-half} & \!\!\!\!\!\!\!\!\!\!\!\text{Spin-zero} \,\, \\
		\\
		\sum_M \epsilon^{M, \mu}_{k_{i} } \epsilon^{M, \nu}_{k_{j}} \equiv \eta^{\mu\nu} \,, \qquad &
		\sum_M \xi^{M, A}_{k_{i} } \bar{u}^{M}_{B,\,k_{j}} \equiv \delta^A_B \,, \qquad &
		1 \times 1  \equiv 1 \, . \qquad 
	\end{matrix}
\end{align}}\noindent
Here $A$ and $B$ are (Dirac) labels for spinors $\xi$ and $\bar{u}$ and for the scalar case we have no sum. Following  \eqref{eq:scat-eq2} and \eqref{eq:delta}, the explicit $\Delta_{p,q}$ parameters are,
\begin{align}\label{Shift0}
&\!\!\!\!\!\!\!\!\!
A_{n-1} (\bm{1}_\sigma, n,\ldots , 4, \bm{k}^M_{23,\sigma} ): \\
&
\Delta_{1,k_{23}}=\frac{m^2+ k_{23}^2 }{2} , \qquad \!\! \!\! \Delta_{1,4}= \frac{ m^2 -k_{23}^2 }{2} ,\,\qquad \ \ \ \ \ \!\!
\Delta_{k_{23},4}= \frac{ -m^2+ k_{23}^2 }{2} , \nonumber\\
&\!\!\!\!\!\!\!\!\!
A_{n-i}(\bm{1}_\sigma,2, \bm{k}^M_{3:i+3,\sigma} ,i+4,\dots, n): \nonumber\\
&
\Delta_{1,2}=\frac{m^2 - k_{3:i+3}^2 }{2} , \quad  \Delta_{1,k_{3:i+3}}= \frac{ m^2 + k_{3:i+3}^2 }{2} ,\quad
\Delta_{2,k_{3:i+3}}= \frac{ -m^2+ k_{3:i+3}^2 }{2} , \nonumber \\
& \!\!\!\!\!\!\!\!\!
A_{i+2}( \bm{k}^M_{i+4:2,\sigma}, i+3,\ldots, 4,\bm{3}_\sigma ): \nonumber\\
&
\Delta_{3,4}=\frac{m^2 - k_{i+4:2}^2 }{2} , \quad  \Delta_{3,k_{i+4:2}}= \frac{ m^2 + k_{i+4:2}^2 }{2} ,\quad
\Delta_{4,k_{i+4:2}}= \frac{ -m^2+ k_{i+4:2}^2 }{2} .\nonumber 
\end{align}
 We are now able to calculate the non-longitudinal contributions in \eqref{Shift0}, 
\begin{align}
&
A_{n-1} (\bm{1}_\sigma, n,\ldots , 4, \bm{k}^M_{23,\sigma} )= \nonumber\\
&
\int d\mu_{n-1} {\rm PT}(1, n,\ldots , 4, k^M_{23})\times \!\! \sum_{\gamma\in S_{n-3}} \!\!\! N^\Delta(1_\sigma,\gamma, \epsilon^M_{23,\sigma})
 {\rm PT} (1,\gamma, k^M_{23}) , \label{1example}
\end{align}
where $\epsilon^M_{23,\sigma}$ takes different expressions depending on the spin of the massive particle. Following the conventions in \eqref{Spiners} we have $\epsilon^M_{23,0} \rightarrow1$, $\epsilon^M_{23,1/2} \rightarrow \xi^{M, A}_{k_{23}}$ and $\epsilon^M_{23,1} \rightarrow \epsilon^{M, \mu}_{k_{23}}$.
\\[5pt] 
For the longitudinal contributions we have to consider, 
\begin{equation}
\sum_L  
	\frac{ A_{n-i}(\bm{1}_\sigma,3_\sigma, \bm{k}^L_{24:i+3},i+4,\dots, n) \, A_{i+2}( \bm{k}^L_{i+4:13}, i+3,i+2,\ldots, 4,\bm{2} ) }{ k_{24:i+3}^2 } .
\end{equation}
Since the configuration has $k_1^2=k_3^2=m^2$ then, 
the $\Delta_{a,b}$ parameters are,
\begin{align}
&\!\!\!\!\!\!\!\!\!\!\!\!
A_{i\!+2}( \bm{k}^L_{i+4:13}, i\!+\!3,i\!+\!2,.., 4,\bm{2} ): \nonumber \\
&
\Delta_{2,4}\!=\!\frac{- k_{i+4:13}^2 }{2} , \ \ \ \ \ \ \ \ \ \ \, \Delta_{2,k_{i+4:13}}\!=\! \frac{ k_{i+4:13}^2 }{2},\ \ \ \ 
\Delta_{4,k_{i+4:13}}\!=\! \frac{ k_{i+4:13}^2 }{2}, \label{Shift1} \\
&\!\!\!\!\!\!\!\!\!\!\!\!
A_{n-i}(\bm{1}_\sigma,3_\sigma, \bm{k}^L_{24:i+3},i+4,\dots, n): \nonumber \\
&
\Delta_{1,3}=\frac{2m^2- k_{24:i+3}^2 }{2} , \quad  \Delta_{1,k_{24:i+3}}= \frac{ k_{24:i+3}^2 }{2} ,\quad
\Delta_{3,k_{24:i+3}}= \frac{ k_{24:i+3}^2 }{2} , \label{Shift2}
\end{align}
with $\sum_L \epsilon^{L,\mu}_i \epsilon^{L,\nu}_j$ as in \eqref{gluingR}. Since the spinning particles do not appear in $A_{i\!+2}( \bm{k}^L_{i+4:13}, i\!+\!3,i\!+\!2,.., 4,\bm{2} )$, we can use the same prescription given in \eqref{1example} with $\sigma=1$, namely, 
\begin{align}\label{2example}
&
A_{i\!+2}( \bm{k}^L_{i+4:13}, i\!+\!3,.., 4,\bm{2} ) = \nonumber \\
&
\int d\mu_{i+2} {\rm PT}({k}_{i+4:13},.., 4,{2})\, \left[ \sum_{\gamma\in S_{i}} N^\Delta(\epsilon^L_{i+4:13} ,\gamma, 2) {\rm PT}(k_{i+4:13},\gamma, 2) \right]\bigg|_{\epsilon^L_{i+4:13}=k_{i+4:13}} .
\end{align}
For $A_{n-i}(\bm{1}_\sigma,3_\sigma, \bm{k}^L_{24:i+3},i+4,\dots, n)$ with $\sigma=1$, we can use the same prescription as in \eqref{2example} and write, 
\begin{align}\label{3example}
&
A_{n-i} (3, \bm{k}^L_{24:i+3},i+4,\dots, n,\bm{1}) = \nonumber \\
&
\int d\mu_{n-i} {\rm PT}(3, {k}_{24:i+3},\dots, n,1) \! \left[ \sum_{\gamma\in S_{n-i-2}} \!\!\!\!\! N^\Delta(\epsilon^L_{24:i+3} ,\gamma, 1) {\rm PT}(k_{24:i+3},\gamma, 1) \right]\bigg|_{\epsilon^L_{24:i+3}=k_{24:i+3}} . 
\end{align}
However when $\sigma=0$ or $\sigma=1/2$ the color-kinematic numerator, $ N^\Delta(\epsilon^L_{24:i+3} ,\gamma, 1)$ needs to be generalized. In order to extend \eqref{3example} for non-integer spinning particles, we propose the following 
\begin{align}\label{4example}
&
A_{n-i} (\bm{3}_\sigma, {k}^L_{24:i+3},i+4,\dots, n,\bm{1}_\sigma) = \nonumber \\
&
\int d\mu_{n-i} {\rm PT}(3, {k}_{24:i+3},\dots, n,1) \! \left[ \sum_{\rho\in S_{n-i-2}} \!\!\!\!\! 
N^\Delta(3_\sigma ,\rho, 1_\sigma) {\rm PT}(3,\rho, 1) \right]\bigg|_{\epsilon^L_{24:i+3}=k_{24:i+3}} ,
\end{align}
where the numerators, $N^\Delta(3_\sigma ,\rho, 1_\sigma)$, are defined by using a generalized field-strength, $f^{\mu\nu}_{24:i+3} \equiv k^\mu_{24:i+3} \epsilon^\nu_{24:i+3} - k^\nu_{24:i+3} \epsilon^\mu_{24:i+3}$. Notice that $f^{\mu\nu}_{24:i+3} = -f^{\nu\mu}_{24:i+3}$ and $(k_{24:i+3} \cdot f_{24:i+3})^{\nu} =0$.  When $\sigma=1$, the computation \eqref{4example} is in agreement with \eqref{3example}, provided that
\begin{equation}\label{LongitudinalI}
	A_{n-i}(3, \bm{k}_{24:i+3},i+4,\dots, n,\bm{1}) =
	A_{n-i}(\bm{3}, k_{24:i+3},i+4,\dots, n,\bm{1}).
\end{equation}
To summerize the analysis, the following generalization of the formula \eqref{Gen-C} is suggested, using $	A_{n-i}(\bm{3}, k^L_{24:i+3},i\!+\!4,.., n,\bm{1}) = 
	(-1)^{n-i-1}\! A_{n-i}(\bm{1},n,..,i+4, k^L_{24:i+3},\bm{3}):$
\begin{align}\label{Gen-C2}
	&
	A_n ( {1}_\sigma,2,{3}_\sigma,\ldots , n)   
	\equiv \int d\mu_n\, {\rm PT}(1,2,\ldots, n)\times \sum_{\gamma\in S_{n-2}} N(1_\sigma,\gamma,3_\sigma) \, {\rm PT}(1,\gamma,3)
	\nonumber\\
	&
	=(-1) \sum_M \frac{ A_{n-1} (\bm{1}_\sigma, n,\ldots , 4, \bm{k}^M_{23,\sigma} ) \, A_{3}( \bm{k}_{4:1,\sigma}^M,2,\bm{3}_\sigma) }{ k_{4:1}^2 -m^2 }  
	-\sum_{i=1}^{n-3} 
	\left( \delta_{n, 2 \mathbb{N}} + (-)^i \, \delta_{n, 2 \mathbb{N}+1} \right)
	\nonumber \\
	&
	\left[
	\sum_M 
	\frac{A_{n-i}(\bm{1}_\sigma,2, \bm{k}^M_{3:i+3,\sigma} ,i+4,\dots, n) \, A_{i+2}( \bm{k}^M_{i+4:2,\sigma}, i+3,\ldots, 4,\bm{3}_\sigma ) }{ k_{3:i+3}^2 -m^2 }  
	\right.
	 \\
	&
	+
	2(-1)^{n-i}
	\left. 
	\sum_L  
	\frac{ A_{n-i}(\bm{1}_\sigma,n, \ldots ,i+4, k^L_{24:i+3},\bm{3}_\sigma)  \, A_{i+2}( \bm{k}^L_{i+4:13}, i+3,i+2,\ldots, 4,\bm{2} ) }{ k_{24:i+3}^2 } 
	\right] ,
	\nonumber
\end{align}
where the massive external particles are, $k_1^2=k_3^2=m^2$, $\sigma$ denotes the spin and the $\Delta_{a,b}$ parameters are given by \eqref{Shift0}, \eqref{Shift1} and \eqref{Shift2}.  We have verified the validity of this construction in appendix \ref{Examples}, were we provide explicit examples at four- five- and six-point. 
\section{Conclusion}\label{conclusion}
We have in this paper presented a new computational tool for numerators with spinning massive particles that require only $(n-2)!$ diagrams. We have shown through several explicit examples and analytic computations how to write compact expressions for numerators in terms of exponential numerators. An interesting feature of this construction is its universal formulation, which implies that we can use the diagrammatic rules regardless of having integer or non-integer spinning massive particles. Direct applications could for instance be in the context of refs.\cite{apply}.\\[5pt]
We have also considered how the formulation of the new exponential numerators for amplitudes allows the extension of recurrence relations for non-integer spins.
We envision applications of our results in the context of computation of gravitational scattering where more compact expressions for tree amplitudes are useful for deriving post-Minkowskian physics in general relativity through unitarity of loop expansions. We note that since amplitudes derived from the diagrammatic scattering equations formalism are not derived from on-shell recursion in four-dimensional space-time, we avoid the appearance of spurious poles and have amplitude expressions which are directly employable for computations in arbitrary dimensions.  \\[5pt]
An interesting point is if it is possible to extend the formalism considered in this paper beyond fundamental spins, for instance, massive objects with large classical spins. Preliminary observations suggest that this will require a further extension of the formalism beyond minimal couplings, and thus we leave it as an interesting task for the future.
\acknowledgments
We thank Johannes Agerskov, Poul H. Damgaard, Oliver Schlotterer and Fei Teng for discussions. We are also very grateful to Y. Geyer for providing us with
her Mathematica package to compute color-kinematic numerators. H.G. acknowledges partial support from University Santiago de Cali (USC).\\[5pt]
\appendix
\section{Recurrence formula at four, five and six points}\label{Examples}
Let us consider the four-point ordered amplitude, $A_4({1}_{1/2} , 2, {4}_{1/2},3)$, where legs one and four are two massive spinors, satisfying $k_1^2=k_4^2=m^2$, and where legs two and three are massless gluons.
These massive fermions satisfy the equations of motion,
\begin{align}
	\bar{u}_1 (\ks_1-m) =0, \quad (\ks_4+m) v_4 =0 \, \Rightarrow \, v_4= (\ks_4-m) \xi_4.
\end{align}
Now, applying the new factorization formula proposed in \eqref{Gen-C2}, we arrive at
\begin{align}\label{}
	&A_4({1}_{1/2} , 2, {4}_{1/2},3)  
	= \nonumber\\
	&
	- 
	\sum_M \frac{ A_{3} ( \bm{1}_{1/2}, 3, \bm{k}^M_{24,{1/2}} ) \, A_{3}( \bm{k}_{31,{1/2}}^M ,2, \bm{4}_{1/2} ) }{ k_{13}^2 -m^2 } -
	\sum_M 
	\frac{A_{3}(\bm{1}_{1/2} ,2, \bm{k}^M_{34,{1/2}} ) \, A_{3}( \bm{k}^M_{12,{1/2} } , 3 , \bm{4}_{1/2} ) }{ k_{34}^2 -m^2 }  
	\nonumber \\
	&
	+
	2\,
	\sum_L  
	\frac{ A_{3}( \bm{1}_{1/2} , {k}^L_{23} , \bm{4}_{1/2} ) \, A_{3}( \bm{k}^L_{14}, 3,\bm{k}_2 ) }{ k_{23}^2 } .
	\qquad
\end{align}
From the three-point color-kinematics numerator for spin-one and spin-1/2 we have,
\begin{align}
	N({a},b,{c}) &= 
	\left( \epsilon_a \cdot \exp[ -\frac{ f_b }{(\epsilon_b\cdot k_a) }] \cdot   \epsilon_c \right) (\epsilon_b\cdot k_a), \\
	N({a}_{1/2},b,{c}_{1/2}) &= 
	\left( \bar{ u}_a \exp[ -\frac{ \f_b }{(\epsilon_b\cdot k_a) }]   \xi_c \right) (\epsilon_b\cdot k_a).
\end{align}
So using \eqref{gluingR} and \eqref{G-glue2} we get
\begin{align}
	& \!\!\!\!\!\!\!\!\!\!\!\!\!\!\!\!\!\!\!\!
	\sum_M  A_{3} ( \bm{1}_{1/2} , 3, \bm{k}^M_{24,{1/2} } ) \, A_{3}( \bm{k}_{31,{1/2}}^M ,2,\bm{4}_{1/2} )  
	=
	\sum_M N({1}_{1/2},3,{\xi}^M_{24}) \times N( {\bar u}^M_{31} ,2,{4}_{1/2})
	\nonumber
	\\ 
	&= 
	\left( \bar{ u}_1  \exp[ - \frac{ \f_3 }{(\epsilon_3\cdot k_1) } ] \exp[ -\frac{ \f_2 }{(\epsilon_2\cdot k_{13} ) } ]  \xi_4 \right) (\epsilon_3\cdot k_1) (\epsilon_2\cdot k_{13} )  ,
\end{align}
\begin{align}
	& \!\!\!\!\!\!\!\!\!\!\!\!\!\!\!\!\!\!\!\!
	\sum_M 
	A_{3}(\bm{1}_{1/2} ,2, \bm{k}^M_{34,{1/2}} ) \, A_{3}( \bm{k}^M_{12,{1/2}} , 3 , \bm{4}_{1/2} )  
	=
	\sum_M N({1}_{1/2},2,{\xi}^M_{34}) \times N( {\bar u}^M_{12} ,3,{4}_{1/2})
	\nonumber\\
	&
	= 
	\left(\overline{ u}_1 \exp [-\frac{ f_2}{ (\epsilon_2\cdot k_1) } ] \exp[- \frac{ \f_3 }{2(\epsilon_3\cdot k_{12} ) }]  \xi_4\right) (\epsilon_2\cdot k_1) (\epsilon_3\cdot k_{12} ) ,
\end{align}
\begin{align}
	\sum_L  
	\frac{ A_{3}(  \bm{1}_{1/2} , {k}^L_{23} , \bm{4}_{1/2} ) \, A_{3}( \bm{k}^L_{14}, 3,\bm{k}_2 ) }{ k_{23}^2 } 
	= 
	\sum_L
	\frac{ N({1}_{1/2},\epsilon^L_{23} , {4}_{1/2} ) N( {\epsilon}^L_{14} ,3,{2})  }{k_{23}^2}
	=
	\frac{  ( \overline{ u}_1 \xi_4 ) (\epsilon_2\cdot \epsilon_3) }{ 4} .
\end{align}
Thus we arrive at
\begin{align}\label{}
	A_4({1}_{1/2} , 2, {4}_{1/2},3) = -\frac{N({1}_{1/2},2,3,{4}_{1/2} ) }{2 (k_1\cdot k_{2})} -  \frac{N({1}_{1/2},3,2,{4}_{1/2})}{2 (k_1\cdot k_{3})} .
\end{align}
where the $N({1}_{1/2},b,c,{4}_{1/2} ) $ is the four-point color-kinematics master numerator,
\begin{align}
	N({1}_{1/2},b,c,{4}_{1/2} )  
	&= 
	\left(\overline{ u}_1  \exp[ -\frac{ \f_b }{(\epsilon_b\cdot k_1) }] \exp[ -\frac{ \f_c }{(\epsilon_c\cdot k_{1b} ) }  ]  \xi_4\right) (\epsilon_b\cdot k_1) (\epsilon_c\cdot k_{1b} )  \nonumber \\
	& -
	\frac{ 1 }{ 2}  ( \overline{ u}_1 \xi_4 ) (\epsilon_b\cdot \epsilon_c) (k_1\cdot k_b) .
\end{align}
By the monodromy identity (on the support of the scattering equations), 
\begin{equation}
	(k_3\cdot k_2) {\rm PT} (1,2,3,4) +(k_3\cdot k_{24}) {\rm PT} (1,2,4,3)=0,  
\end{equation}
it is straightforward to check that,
\begin{align}\label{}
	A_4({1}_{1/2} , 2, 3, {4}_{1/2}) &= -\frac{ (k_3 \cdot k_{24}) }{ (k_3\cdot k_2) } \, A_4({1}_{1/2} , 2, {4}_{1/2} ,3 ) \nonumber\\
	&=
	\frac{N({1}_{1/2},2,3,{4}_{1/2} ) }{2 (k_1\cdot k_{2})} +  \frac{ N({1}_{1/2},2,3,{4}_{1/2} ) - N({1}_{1/2},3,2,{4}_{1/2})}{2 (k_2\cdot k_{3})} 
	.
\end{align}
We note that the color-kinematics numerator $N({1}_\sigma,2,3,{4}_\sigma )$ ($N( {1}_\sigma ,3,2,{4}_\sigma$) is factorized as
\begin{align}\label{N4p-F}
	&
	N({1}_\sigma,2,3,{4}_\sigma )= \nonumber\\
	&
	\sum_M N( {1}_\sigma , 2, {\epsilon}^M_{34,\sigma} ) \, N( {\epsilon}^M_{12,\sigma} , 3 , {4}_\sigma ) - \frac{1}{2} \,
	\frac{(k_1\cdot k_2)}{ (k_{2}\cdot k_3) } \sum_L  
	N( {1}_\sigma , \epsilon^L_{23} , {4}_\sigma ) \, N ( {\epsilon}^L_{14}, 3,{2} ) ,
\end{align}
where $\sigma$ is denoting spin. An interesting observation is that we can understand this equation directly in terms of the graphical rules.\\[5pt]
For the five-point amplitude, we take particles one and five to be massive spin-1/2 particles ($k_1^2=k_5^2=m^2$) and while particles two, three, and four are gluons. From the recurrence formula \eqref{Gen-C2}, we have
\begin{align}\label{}
	&
	A_5({1}_{1/2},2,{5}_{1/2},3,4)
	=  \nonumber\\
	&
	- \sum_M \frac{ A_{4} ( \bm{1}_{1/2}, 4,3, \bm{k}^M_{25,{1/2}} ) \, A_{3}( \bm{k}_{341,{1/2}}^M,2,\bm{5}_{1/2} ) }{ k_{25}^2- m^2 }  
	- \sum_M \frac{ A_{3} (\bm{1}_{1/2}, 2, \bm{k}^M_{345,{1/2}} ) \, A_{4}( \bm{k}_{12,{1/2}}^M,4,3,\bm{5}_{1/2} ) }{ k_{12}^2- m^2 }  
	\nonumber \\
	&
	+ \sum_M \frac{ A_{4} (\bm{1}_{1/2}, 2, \bm{k}^M_{35,{1/2} },4 ) \, A_{3}( \bm{k}_{124,{1/2}}^M,3,\bm{5}_{1/2} ) }{ k_{35}^2- m^2 }  
	+
	2\,
	\sum_L  
	\frac{ A_{3}(\bm{1}_{1/2} , {k}^L_{234} , \bm{5}_{1/2}) \, A_{4}( \bm{k}^L_{15},4, 3,\bm{2} ) }{ k_{234}^2 } 
	\nonumber \\
	&
	+
	2\,
	\sum_L  
	\frac{ A_{4}(\bm{1}_{1/2},4, {k}^L_{23}, \bm{5}_{1/2}) \, A_{3}( \bm{k}^L_{145}, 3,\bm{2} ) }{ k_{23}^2 } .
\end{align}
Using the parameters $\Delta_{ab}$  
and \begin{align}
	N({a}_{1/2},b,c,{d}_{1/2} ) &= 
	\left(\overline{ u}_a  \exp[ -\frac{ \f_b }{(\epsilon_b\cdot k_a) }] \exp[ -\frac{ \f_c }{(\epsilon_c\cdot k_{ab} ) }  ]  \xi_d\right) (\epsilon_b\cdot k_a) (\epsilon_c\cdot k_{ab} ) \nonumber\\
	&
	-
	\frac{ 1 }{ 2}  ( \overline{ u}_a \xi_d ) (\epsilon_b\cdot \epsilon_c) (k_a\cdot k_b+\Delta_{ab}) ,
\end{align} we
arrive at
\begin{align}
	&\!\!\!\!\!\!\!\!\!\!\!\!\!
	A_{4} (\bm{1}_{1/2}, 4,3, \bm{k}^M_{25,{1/2}} ) A_{3}( \bm{k}_{341,{1/2}}^M,2,\bm{5}_{1/2} ) = \nonumber\\
	& 
	\left[
	\frac{N({1}_{1/2},4,3,{\xi}^M_{25} ) }{2 (k_4\cdot k_{1})} +  \frac{ N({1}_{1/2},4,3,{\xi}^M_{25} ) - N({1}_{1/2},3,4,{\xi}^M_{25} ) ) }{2 (k_4\cdot k_{3})} \right]
	N( {\bar u}^M_{341},2,{5}_{1/2} ) ,\nonumber \\
	&
	\!\!\!\!\!\!\!\!\!\!\!\!\!	\text{with}, \, \,\,\,\, 
	\Delta_{13}=\frac{m^2-k_{25}^2}{2}, \quad \Delta_{1\,k_{25}}=\frac{m^2+k_{25}^2}{2}, \quad \Delta_{3\,k_{25}}=\frac{-m^2+k_{25}^2}{2} . \nonumber
\end{align}
\begin{align}
	&\!\!\!\!\!\!\!\!\!\!\!\!\!
	A_{3} (\bm{1}_{1/2}, 2, \bm{k}^M_{345,{1/2} } ) A_{4}( \bm{k}_{12,{1/2}}^M,4,3,\bm{5}_{1/2} ) =\nonumber\\
	& 
	N( {1}_{1/2},2, {\xi}^M_{345} )
	\left[
	\frac{N({\bar u}^M_{12},4,3,{5}_{1/2} ) }{2 (k_4\cdot k_{12})} +  \frac{ N({\bar u}^M_{12},4,3,{5}_{1/2} ) - N({\bar u}^M_{12},3,4,{5}_{1/2} ) }{2 (k_4\cdot k_{3})} \right] ,
	\nonumber \\
	&
	\!\!\!\!\!\!\!\!\!\!\!\!\!	\text{with}, \, \,\,\,\, 
	\Delta_{k_{12}\,3}=\frac{k_{12}^2-m^2}{2}, \quad \Delta_{k_{12}\,5}=\frac{k_{12}^2 + m^2}{2}, \quad \Delta_{35}=\frac{-k_{12}^2+m^2}{2} .\nonumber
\end{align}
\begin{align}
	&
	\!\!\!\!\!\!\!\!\!\!\!\!\!\!\!\!\!\!\!\!\!\!\!\!\!\!\!\!\!\!\!\!\!\!\!\!	A_{4} (\bm{1}_{1/2}, 2, \bm{k}^M_{35,{1/2} },4 ) A_{3}( \bm{k}_{124,{1/2}}^M,3,\bm{5}_{1/2} )
	= \nonumber\\
	&
	\left[-
	\frac{N({1}_{1/2},2,4, {\xi}^M_{35} ) }{2 (k_4\cdot k_{35})} -  \frac{ N({1}_{1/2},4,2, {\xi}^M_{35} ) }{2 (k_4\cdot k_{1})} \right]
	N( {\bar u}^M_{124} ,3, {5}_{1/2} )
	,
	\nonumber \\
	&
	\!\!\!\!\!\!\!\!\!\!\!\!\!\!\!\!\!\!\!\!\!\!\!\!\!\!\!\!\!\!\!\!\!\!\!	\text{with}, \, \,\,\,\, 
	\Delta_{12}=\frac{m^2-k_{35}^2}{2}, \quad \Delta_{1\,k_{35}}=\frac{m^2+k_{35}^2}{2}, \quad \Delta_{2\,k_{35}}=\frac{-m^2+k_{35}^2}{2} . \nonumber
\end{align}
\begin{align}
	&
	\!\!\!\!\!\!\!\!\!\!\!\!\!\!\!\!\!\!\!\!	\!\! A_{3}(\bm{1}_{1/2} , {k}^L_{234} , \bm{5}_{1/2}) \, A_{4}( \bm{k}^L_{15},4, 3,\bm{2} ) 
	= \nonumber\\
	&
	N( {1}_{1/2},\epsilon_{234}^L, {5}_{1/2} )
	\left[
	\frac{N({\epsilon}^L_{15},4,3, {2} ) }{2 (k_4\cdot k_{15})} +  \frac{ N({\epsilon}^L_{15},4,3,{2} ) - N({\epsilon}^L_{15},3,4,{2} )}{2 (k_4\cdot k_{3})} \right] ,
	\nonumber \\
	&\!\!\!\!\!\!\!\!\!\!\!\!\!\!\!\!\!\!\!\!\!
	\text{with}, \, \,\,\,\, 
	\Delta_{k_{15}\,2}=\frac{k_{15}^2}{2}, \quad \Delta_{k_{15}\,3}=\frac{k_{15}^2}{2}, \quad \Delta_{23}=\frac{-k_{15}^2}{2} .\nonumber
\end{align}
\begin{align}
	& \!\!\!\!\!\!\!\!\!
	A_{4}(\bm{1}_{{1/2}},4, {k}^L_{23}, \bm{5}_{{1/2}}) \, A_{3}( \bm{k}^L_{145}, 3,\bm{2} ) 
	= \nonumber\\
	&
	\left[
	\frac{ N({1}_{1/2},4,\epsilon^L_{23},{5}_{1/2} ) }{2 (k_4\cdot k_{1})} +  \frac{ N({1}_{1/2},4,\epsilon^L_{23},{5}_{1/2} )  - N({1}_{1/2},\epsilon^L_{23},4,{5}_{1/2} ) }{2 (k_4\cdot k_{23})} \right] 
	N( {\epsilon}_{145}^L, 3 , {2} ) 
	,
	\nonumber \\
	& \!\!\!\!\!\!\!\!\!
	\text{with}, \, \,\,\,\, 
	\Delta_{1\,k_{23}}=\frac{k_{23}^2}{2}, \quad \Delta_{15}=\frac{2m^2-k_{23}^2}{2}, \quad \Delta_{k_{23}\,5}=\frac{k_{23}^2}{2} .\nonumber
\end{align}
By relabelling, the same method can be used to compute, $A_5({1}_{1/2},4,{5}_{1/2},3,2)$. Therefore, from the monodromy relationship (on the support of the scattering equations)
\begin{equation}
	(k_5\cdot k_2) {\rm PT} (2,5,3,4,1) +(k_5\cdot k_{23}) {\rm PT} (2,3,5,4,1)+(k_5\cdot k_{234}) {\rm PT} (2,3,4,5,1)=0, 
	\nonumber 
\end{equation}
the amplitude, $ A_5({1}_{1/2},2,3,4,{5}_{1/2}) $, becomes
\begin{equation}
	A_5({1}_{1/2},2,3,4,{5}_{1/2}) = -\frac{(k_5\cdot k_2)}{ (k_5\cdot k_{234} )} A_5({1}_{1/2},2,{5}_{1/2},3,4) + 
	\frac{(k_5\cdot k_{23})}{ (k_5\cdot k_{234} )} A_5({1}_{1/2},4,{5}_{1/2},3,2). \nonumber
\end{equation}
\\[5pt]
As a final example, the six-point amplitude, $A_6({1}_{1/2},2,{6}_{1/2},3,4,5)$, where one and six are massive spin-1/2 particles ($k_1^2=k_6^2=m^2$), turns into,  
\begin{align}\label{}
	&
	A_6({1}_{1/2},2,{6}_{1/2},3,4,5)
	= (-1) \times \sum_M \left[ \frac{ A_{5} ( \bm{1}_{1/2}, 5,4,3, \bm{k}^M_{26,{1/2}} ) \, A_{3}( \bm{k}_{3451,{1/2}}^M,2,\bm{6}_{1/2} ) }{ k_{26}^2- m^2 } + \right. \nonumber\\
	&
	\frac{ A_{4} (\bm{1}_{1/2}, 2, \bm{k}^M_{346,{1/2}},5 ) \, A_{4}( \bm{k}_{512,{1/2}}^M,4,3,\bm{6}_{1/2} ) }{ k_{346}^2- m^2 } 
	+ \frac{ A_{3} (\bm{1}_{1/2}, 2, \bm{k}^M_{3456,{1/2}} ) \, A_{5}( \bm{k}_{12,{1/2}}^M,5,4,3,\bm{6}_{1/2} ) }{ k_{12}^2- m^2 } +
	\nonumber \\
	&
	\left.
	\frac{ A_{5} (\bm{1}_{1/2}, 2, \bm{k}^M_{36,{1/2}},4,5 ) \, A_{3}( \bm{k}_{4512,{1/2}}^M,3,\bm{6}_{1/2} ) }{ k_{36}^2- m^2 } \right] 
	+
	2
	\sum_L\left[  
	\frac{ A_{3}(\bm{1}_{1/2} , {k}^L_{2345} , \bm{6}_{1/2}) \, A_{5}( \bm{k}^L_{16},5,4, 3,\bm{2} ) }{ k_{2345}^2 } \right.
	\nonumber \\
	& \left.
	-
	\frac{ A_{4}(\bm{1}_{1/2} , 5, {k}^L_{234} , \bm{6}_{1/2}) \, A_{4}( \bm{k}^L_{561},4, 3,\bm{2} ) }{ k_{234}^2 } 
	+
	\frac{ A_{5}(\bm{1}_{1/2} , 5,4, {k}^L_{23} , \bm{6}_{1/2}) \, A_{3}( \bm{k}^L_{4561}, 3,\bm{2} ) }{ k_{23}^2 } \right].
\end{align}
We have checked all results in this appendix numerically.
\section{Diagrams contributing at six points}\label{app:num1}
Below we have listed the 24 diagrams contributing at six points. 
\begin{align}
\begin{BCJ}[6]
\baseline{2}
\baseline{3}
\baseline{4}
\baseline{5}
\end{BCJ}
=&(\epsilon_2\cdot k_{1})( \epsilon_3 \cdot k_{12}) (\epsilon_4 \cdot k_{123})(\epsilon_5 \cdot k_{1234})
\\ 
&\!\!\!\!\!\!\!\!\!\!\!\!\!\!\!\!\!\!\!\!\!\!\!\!\!\!\!\!\!\!\!\!\!\!\!\!\!\!\!\!\!\!\!\!\!\!\!\!\!\!\times\left(
\epsilon_1\cdot \exp[-\frac{f_2}{\epsilon_{2}\cdot k_1}]\cdot \exp[-\frac{f_3}{\epsilon_3 \cdot k_{12}}]\right. \left.\cdot \exp[-\frac{f_4}{\epsilon_4 \cdot k_{123}}]\cdot \exp[-\frac{f_5}{\epsilon_5 \cdot k_{1234}}]\cdot\epsilon_6 
\right)  \nonumber
\\
\begin{BCJ}[6]
\baseline{2}
\baseline{3}
\one{4}
\one{5}
\num{4}{5}{1}
\finaltwo{4}{3}
\end{BCJ}=& \frac{(-1)}{2}\black{\left(k_4\cdot k_{123}\right)} \black{(\epsilon _4\cdot \epsilon _5)}(\epsilon _2\cdot k_1)\left(\epsilon_3\cdot k_{12}\right)\\
&\times\left(
\epsilon_1\cdot \exp[-\frac{f_2}{\epsilon_{2}\cdot k_1}]\cdot \exp[-\frac{f_3}{\epsilon_3 \cdot k_{12}}]\cdot\epsilon_6 
\right) \nonumber
\\
\begin{BCJ}[6]
\baseline{2}
\baseline{4}
\one{3}
\one{5}
\num{3}{5}{1}
\finaltwo{3}{2}
\end{BCJ}
=&\frac{(-1)}{2}\black{\left(k_3\cdot k_{12}\right)}\black{(\epsilon_3\cdot \epsilon_5)}(\epsilon_{2}\cdot k_1)(\epsilon_{4}\cdot k_{12})
\\  \nonumber 
&\times
\left(
\epsilon_1\cdot \exp[-\frac{f_2}{\epsilon_{2}\cdot k_1}]\cdot  \exp[-\frac{f_4}{\epsilon_{4}\cdot k_{12}}]\cdot \epsilon_6
\right)\\
\begin{BCJ}[6]
\baseline{2}
\baseline{5}
\one{3}
\one{4}
\num{3}{4}{1}
\finaltwo{3}{2}
\end{BCJ}
=&\frac{(-1)}{2}\black{\left(k_3\cdot k_{12}\right)}
\black{(\epsilon _3\cdot \epsilon _4)}(\epsilon _2\cdot k_1)\left(\epsilon _5\cdot k_{12}\right)\\  \nonumber
&\times
\left(
\epsilon_1\cdot \exp[-\frac{f_2}{\epsilon_{2}\cdot k_1}]\cdot \exp[-\frac{f_5}{\epsilon_5 \cdot k_{12}}]\cdot\epsilon_6 
\right)
\end{align}
\begin{align}
\begin{BCJ}[6]
\baseline{3}
\baseline{4}
\one{2}
\one{5}
\num{2}{5}{1}
\finalone{2}
\end{BCJ}
=&\frac{(-1)}{2}\black{(k_1\cdot k_2)}\black{( \epsilon _2\cdot \epsilon _5 )}(\epsilon_{3}\cdot k_{1})\\  \nonumber
&\times\left(\epsilon _4\cdot k_{13}\right)
\left(
\epsilon_1\cdot \exp[-\frac{f_3}{\epsilon_{3}\cdot k_{1}}]\cdot  \exp[-\frac{f_4}{\epsilon_{4}\cdot k_{13}}]\cdot \epsilon_6
\right) \\
\begin{BCJ}[6]
\baseline{3}
\baseline{5}
\one{2}
\one{4}
\num{2}{4}{1}
\finalone{2}
\end{BCJ}
=&\frac{(-1)}{2}\black{(k_1\cdot k_2)}\black{(\epsilon_2\cdot\epsilon_4)}(\epsilon_3\cdot k_1)(\epsilon_5\cdot k_{13})\\  \nonumber
&\times\left(
\epsilon_1\cdot \exp[-\frac{f_3}{\epsilon_{3}\cdot k_{1}}]\cdot  \exp[-\frac{f_5}{\epsilon_{5}\cdot k_{13}}]\cdot \epsilon_6\right)\\
\begin{BCJ}[6]
\baseline{4}
\baseline{5}
\one{2}
\one{3}
\num{2}{3}{1}
\finalone{2}
\end{BCJ}
=&\frac{(-1)}{2}\black{(k_1\cdot k_2)}\black{(\epsilon _2\cdot \epsilon _3)} (\epsilon _4\cdot
k_1)\left(\epsilon _5\cdot k_{14}\right)\\  \nonumber
&\times
\left(
\epsilon_1\cdot \exp[-\frac{f_4}{\epsilon_{4}\cdot k_{1}}]\cdot  \exp[-\frac{f_5}{\epsilon_{5}\cdot k_{14}}]\cdot \epsilon_6
\right)
\\
\begin{BCJ}[6]
\baseline{2}
\one{3}
\one{4}
\one{5}
\num{3}{5}{1}
\finaltwo{3}{2}
\end{BCJ}
=&\frac{1}{3}\black{(k_3\cdot k_{12})}\black{ N(3,4,5)}(\epsilon_{2}\cdot k_1)
\left(
\epsilon_1\cdot \exp[-\frac{f_2}{\epsilon_{2}\cdot k_1}]\cdot \epsilon_6
\right)\\
\begin{BCJ}[6]
\baseline{3}
\one{2}
\one{4}
\one{5}
\num{2}{5}{1}
\finalone{2}
\end{BCJ}
=&\frac{1}{3}\black{(k_{1}\cdot k_2) }\black{N(2,4,5)}(\epsilon_{3}\cdot k_1)
\left(
\epsilon_1\cdot \exp[-\frac{f_3}{\epsilon_{3}\cdot k_1}]\cdot \epsilon_6
\right)\\
\begin{BCJ}[6]
\baseline{4}
\one{2}
\one{3}
\one{5}
\num{2}{5}{1}
\finalone{2}
\end{BCJ}
=&\frac{1}{3}\black{(k_{1}\cdot k_2)}\black{N(2,3,5)}(\epsilon_{4}\cdot k_1)
\left(
\epsilon_1\cdot \exp[-\frac{f_4}{\epsilon_{4}\cdot k_1}]\cdot \epsilon_6
\right)\\
\begin{BCJ}[6]
\baseline{5}
\one{2}
\one{3}
\one{4}
\num{2}{4}{1}
\finalone{2}
\end{BCJ}
=&\frac{1}{3}\black{(k_{1}\cdot k_2)}\black{N(2,3,4)}(\epsilon_{5}\cdot k_{1})
\left(
\epsilon_1\cdot  \exp[-\frac{f_5}{\epsilon_{5}\cdot k_{1}}]\cdot \epsilon_6
\right)
\\
\begin{BCJ}[6]
\baseline{2}
\one{3}
\one{4}
\two{5}
\num{3}{4}{1}
\edgetwo{5}{4}{3}{2}{1}
\finaltwo{3}{2}
\end{BCJ}
=&\frac{(-1)}{3}\black{(k_3\cdot k_{12})}
\black{(\epsilon_3\cdot \epsilon_4)}
\black{(\epsilon_5\cdot k_{34})}
(\epsilon_{2}\cdot k_1)
\left(
\epsilon_1\cdot \exp[-\frac{f_2}{\epsilon_{2}\cdot k_1}]\cdot \epsilon_6
\right)
\\
\begin{BCJ}[6]
\baseline{3}
\one{2}
\one{4}
\two{5}
\num{2}{4}{1}
\edgetwo{5}{4}{2}{2}{1}
\finalone{2}
\end{BCJ}
=&\frac{(-1)}{3}\black{(k_{1}\cdot k_2)}
\black{(\epsilon_2\cdot \epsilon_4)}
\black{(\epsilon_5\cdot k_{24})}
(\epsilon_{3}\cdot k_1)
\left(
\epsilon_1\cdot \exp[-\frac{f_3}{\epsilon_{3}\cdot k_1}]\cdot \epsilon_6
\right)
\\[-10pt]
\begin{BCJ}[6]
\baseline{4}
\one{2}
\one{3}
\two{5}
\num{2}{3}{1}
\edgetwo{5}{3}{2}{2}{1}
\finalone{2}
\end{BCJ}
=&\frac{(-1)}{3}\black{(k_{1}\cdot k_2)}
\black{(\epsilon _2\cdot \epsilon _3)}
\black{\left(\epsilon _5\cdot k_{23}\right)}
(\epsilon_{4}\cdot k_1)
\left(
\epsilon_1\cdot \exp[-\frac{f_4}{\epsilon_{4}\cdot k_1}]\cdot \epsilon_6
\right)
\end{align}
	\begin{align}
		\begin{BCJ}[6]
			\baseline{5}
			\one{2}
			\one{3}
			\two{4}
			\num{2}{3}{1}
			\edgetwo{4}{3}{2}{2}{1}
			\finalone{2}
		\end{BCJ}
		=&\frac{(-1)}{3}\black{(k_{1}\cdot k_2)}
		\black{(\epsilon_2\cdot \epsilon_3)}
		\black{(\epsilon_4\cdot k_{23})}
		(\epsilon_{5}\cdot k_{1})
		\left(
		\epsilon_1\cdot  \exp[-\frac{f_5}{\epsilon_{5}\cdot k_{1}}]\cdot \epsilon_6
		\right)
		\\
		\begin{BCJ}[6]
			\one{2}
			\one{3}
			\one{4}
			\one{5}
			\numtwo{2}{5}{1}
			\finalone{2}
		\end{BCJ}
		=&\frac{(-1)}{4}\black{(k_1\cdot k_2)}\black{N(2,3,4,5)}(\epsilon_1\cdot \epsilon_6)
\\[-7pt]
		\begin{BCJ}[6]
			\one{2}
			\one{3}
			\one{4}
			\two{5}
			\num{2}{4}{1}
			\edgetwo{5}{4}{2}{2}{1}
			\finalone{2}
		\end{BCJ}
		=&\frac{1}{4}\black{(k_1\cdot k_2)}\black{N(2,3,4)}\black{(\epsilon_{5}\cdot k_{234})}(\epsilon_1\cdot \epsilon_6)
\\[-7pt]
		\begin{BCJ}[6]
			\two{2}
			\two{4}
			\one{3}
			\one{5}
			\num{2}{4}{2}
			\num{3}{5}{1}
			\finalthree{2}
			\finalone{3}
		\end{BCJ}
		=&\frac{1}{4}\black{(k_1\cdot k_2) }\black{(\epsilon _2\cdot \epsilon _4)}
		\black{\left(k_1\cdot k_3\right)}\black{(\epsilon _3\cdot \epsilon_5)}
		(\epsilon _1\cdot \epsilon _6 )
\\[-7pt]
		\begin{BCJ}[6]
			\two{2}
			\two{5}
			\one{3}
			\one{4}
			\num{2}{5}{2}
			\num{3}{4}{1}
			\finalthree{2}
			\finalone{3}
		\end{BCJ}
		=&\frac{1}{4}\black{(k_1\cdot k_2) }\black{(\epsilon _2\cdot \epsilon _5)}
		\black{\left(k_1\cdot k_3\right)}\black{(\epsilon _3\cdot \epsilon_4)}
		(\epsilon _1\cdot \epsilon _6 )
\\[-7pt]
		\begin{BCJ}[6]
			\two{2}
			\two{3}
			\one{4}
			\one{5}
			\num{2}{3}{2}
			\num{4}{5}{1}
			\finalthree{2}
			\finalone{4}
		\end{BCJ}
		=&\frac{1}{4}\black{(k_1\cdot k_2) }\black{(\epsilon _2\cdot \epsilon _3)}
		\black{\left(k_1\cdot k_4\right)}\black{(\epsilon _4\cdot \epsilon_5)}(\epsilon _1\cdot \epsilon _6 )
\\[-7pt]
		\begin{BCJ}[6]
			\one{2}
			\one{3}
			\two{4}
			\three{5}
			\num{2}{3}{1}
			\edgetwo{4}{3}{2}{2}{1}
			\edgesingle{5}{4}{3}{2}
			\finalone{2}
		\end{BCJ}
		=&\frac{(-1)}{4}\black{(k_1\cdot k_2)}\black{( \epsilon _2\cdot \epsilon _3 )}\black{(\epsilon_4\cdot k_{23})}\black{(\epsilon_{5}\cdot k_{4})}(\epsilon_1\cdot \epsilon_6)
\\[-7pt]
		\begin{BCJ}[6]
			\one{2}
			\one{3}
			\two{4}
			\three{5}
			\num{2}{3}{1}
			\edgetwo{4}{3}{2}{2}{1}
			\edgesingletwo{5}{3}{2}{3}{1}
			\finalone{2}
		\end{BCJ}
		=&\frac{(-1)}{4}\black{(k_1\cdot k_2)}\black{( \epsilon _2\cdot \epsilon _3 )}\black{(\epsilon_4\cdot k_{23})}\black{(\epsilon_{5}\cdot k_{23})}(\epsilon_1\cdot \epsilon_6)
\\[-7pt]
		\begin{BCJ}[6]
			\one{2}
			\one{4}
			\two{3}
			\two{5}
			\num{2}{4}{1}
			\num{3}{5}{2}
			\edgeone{3}{2}{2}{1}
			\finalone{2}
		\end{BCJ}
		=&
		\frac{1}{8}\black{\left(k_1\cdot
			k_2\right)} \black{(\epsilon _3\cdot \epsilon _5) (\epsilon _2\cdot \epsilon _4)}\black{\left(k_3\cdot k_{2}\right)}( \epsilon _1\cdot
		\epsilon _6 )\\[-10pt]
		\begin{BCJ}[6]
			\one{2}
			\one{3}
			\two{4}
			\two{5}
			\num{2}{3}{1}
			\num{4}{5}{2}
			\edgetwo{4}{3}{2}{2}{1}
			\finalone{2}
		\end{BCJ}
		=&
		\frac{1}{8}\black{\left(k_1\cdot
			k_2\right)} \black{(\epsilon _4\cdot \epsilon _5) (\epsilon _2\cdot \epsilon _3)}\black{\left(k_4\cdot k_{23}\right)}( \epsilon _1\cdot
		\epsilon _6 )
	\end{align}
\newpage
\section{All seven-point diagrams}\label{app:num2}
We mark diagrams containing an off-shell numerator with squares.\newline
\hspace*{-0.4cm}\includegraphics[width=1\linewidth, page=1]{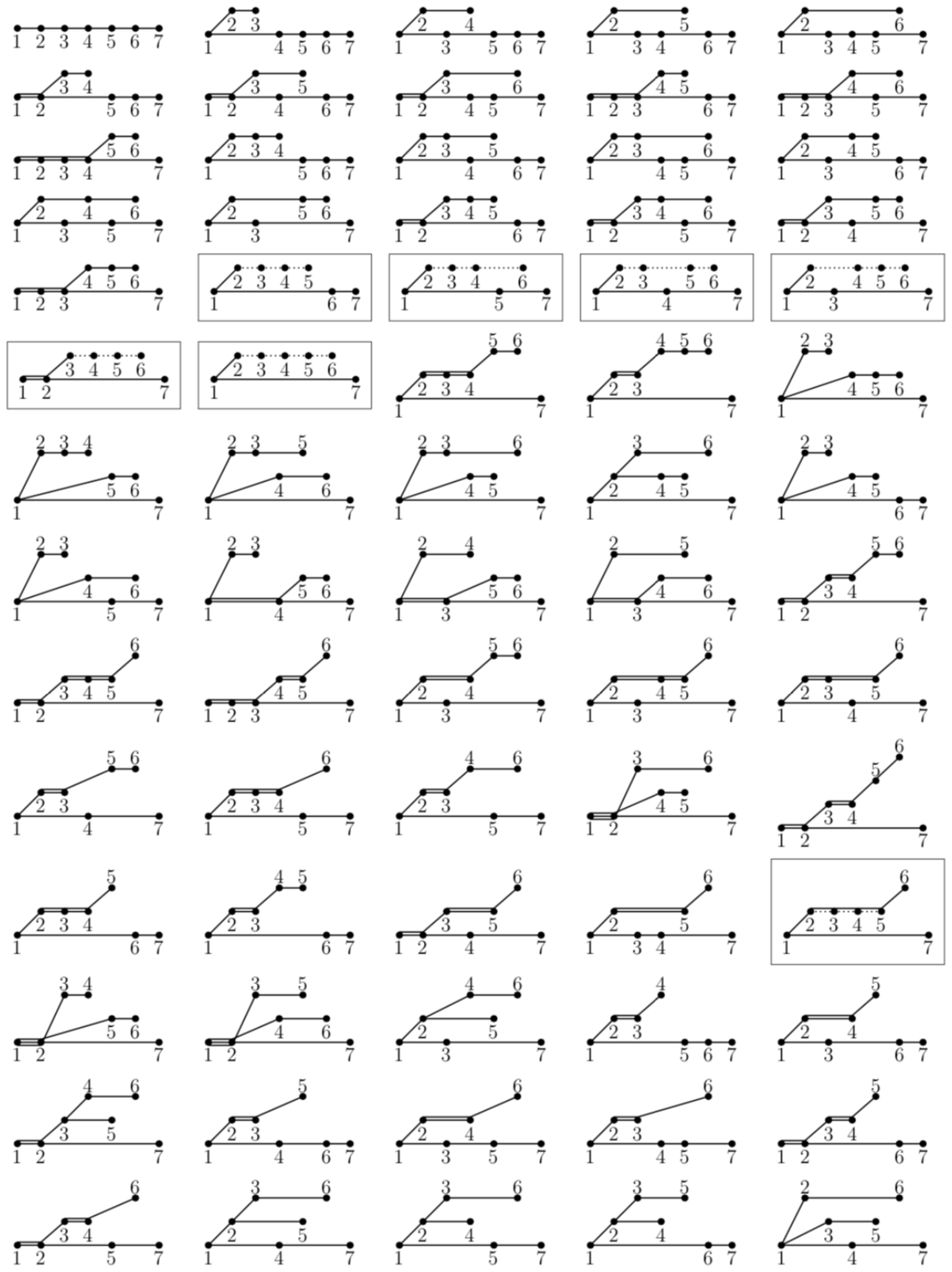}
\newpage
\hspace*{-1cm}\includegraphics[width=1\linewidth, page=1]{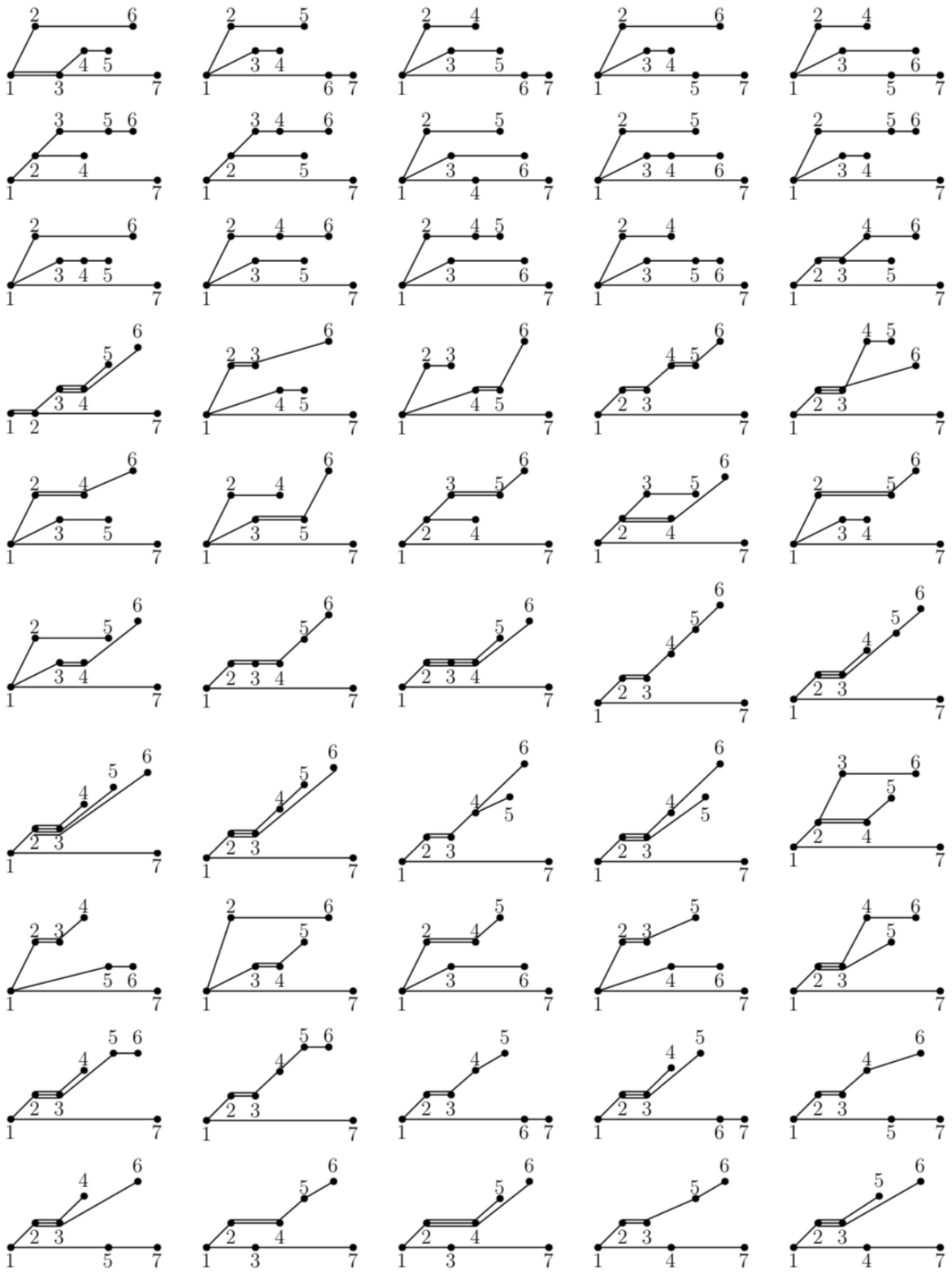}

\end{document}